\documentclass[aps,superscriptaddress,twocolumn,nofootinbib]{revtex4-2}
\usepackage[utf8]{inputenc} % allow utf-8 input
\usepackage[T1]{fontenc}    % use 8-bit T1 fonts
\usepackage[hidelinks]{hyperref}       % hyperlinks
\usepackage{url}            % simple URL typesetting
\usepackage{booktabs}       % professional-quality tables
\usepackage{nicefrac}       % compact symbols for 1/2, etc.
\usepackage{graphicx}
\usepackage{mathtools}
\usepackage{xcolor}
\usepackage{amsmath}
\usepackage{amssymb}
\usepackage{amsfonts}       % blackboard math symbols
\usepackage{physics}
\usepackage{xspace}
\usepackage{bigints}
\usepackage{appendix}
\usepackage{cleveref}
\usepackage{mathtools}
\usepackage{subfig}
\usepackage{paralist}

\newtheorem{theorem}{Theorem}
\newtheorem{definition}{Definition}
\newtheorem{corollary}{Corollary}

\newtheorem{lemma}{Lemma}

\newcommand{\Var}{{\rm Var}}

\newcommand{\bigO}{\ensuremath{\mathcal{O}}\xspace}

\newcommand{\I}{\ensuremath{\mathcal{I}}\xspace}
\newcommand{\vtheta}{\ensuremath{\vec{\theta}}\xspace}

\newcommand{\E}{\ensuremath{\operatorname{\mathbb E}\xspace}}

\newcommand{\epsM}{\ensuremath{\epsilon_{\rm M}}}
\newcommand{\epsS}{\ensuremath{\epsilon_{\rm S}}}
\newcommand{\epsMp}{\ensuremath{\epsM \cdot \sqrt{m}}}
\newcommand{\Beta}{\ensuremath{\operatorname{\mathcal B}}\xspace}
\newcommand{\Obs}{\ensuremath{\hat{O}}}
\newcommand{\Olocal}{\ensuremath{\hat{O}_{\rm Local}}}
\newcommand{\Oglobal}{\ensuremath{\hat{O}_{\rm Global}}}
\newcommand{\Normgrad}{\ensuremath{\norm{\nabla C}}}
\newcommand{\initstate}{\ensuremath{\ket{\psi_0}}}
\newcommand{\sampcost}{\ensuremath{\bar{C}}}
\DeclarePairedDelimiterX{\inp}[2]{\langle}{\rangle}{#1, #2}

\crefname{appendix}{appendix}{appendices}

\graphicspath{ {figures/} }

\newcommand{\aqa}{$\langle aQa ^L\rangle $ Applied Quantum Algorithms, Universiteit Leiden}
\newcommand{\lorentz}{Instituut-Lorentz, Universiteit Leiden, Niels Bohrweg 2, 2333 CA Leiden, Netherlands}
\newcommand{\liacs}{LIACS, Universiteit Leiden, Niels Bohrweg 1, 2333 CA Leiden, Netherlands}

\begin{document}
\title{Analyzing variational quantum landscapes with information content}
\author{Adrián Pérez-Salinas}
\email[]{perezsalinas@lorentz.leidenuniv.nl}
\affiliation{\aqa}
\affiliation{\lorentz}
\author{Hao Wang}
\email[]{h.wang@liacs.leidenuniv.nl}
\affiliation{\aqa}
\affiliation{\liacs}
\author{Xavier Bonet-Monroig}
\email[]{bonet@lorentz.leidenuniv.nl}
\affiliation{\aqa}
\affiliation{\lorentz}

\begin{abstract}
The parameters of the quantum circuit in a variational quantum algorithm induce a landscape that contains the relevant information regarding its optimization hardness.
In this work we investigate such landscapes through the lens of information content, a measure of the variability between points in parameter space.
Our major contribution connects the information content to the average norm of the gradient, for which we provide robust analytical bounds on its estimators.
This result holds for any (classical or quantum) variational landscape.
We validate the analytical understating by numerically studying the scaling of the gradient in an instance of the barren plateau problem.
In such instance we are able to estimate the scaling pre-factors in the gradient.
Our work provides a new way to analyze variational quantum algorithms in a data-driven fashion well-suited for near-term quantum computers. 
\end{abstract}

\maketitle

\section{Introduction}
Variational quantum algorithms (VQAs) have been marked as a promising path towards quantum advantage in pre-fault-tolerant quantum hardware.
In nearly a decade of research since its original proposal~\cite{peruzzo2014variational}, the field of VQAs has seen significant progress both theoretical and experimentally~\cite{cerezo2020variational, bharti2022noisy}.
It is yet to be seen if noisy intermediate-scale quantum (NISQ)~\cite{preskill2018quantum} devices are able to reach unambiguous quantum advantage through VQA.
Issues such as vanishing gradients or barren plateaus (BP)~\cite{mcclean2018barren,cerezo2021cost,wang2021noiseinduced,larocca2022diagnosing}, the expressivity of the quantum circuits~\cite{Herasymenko2021diagrammatic,du2022efficient, holmes2022connecting} or difficulties optimizing a noisy cost function~\cite{bonet-monroig2023performance} are only a few examples of the hurdles faced by VQA which reduce the hope of quantum advantage in the near-term.

From a computer science point-of-view VQAs are a fascinating object of study.
They can be considered classical cost functions with classic input/output.
Yet the cost function might not be classically accessible in general.
So far, there is no clear evidence that optimizing a VQA is feasible with standard optimization methods~\cite{bonet-monroig2023performance}.
Some researchers have attempted to close this gap by developing new optimizers tailored to quantum circuits~\cite{Nakanish2020SOFF,Ostaszewski2021rotosolve} or use machine learning techniques to assist during the optimization~\cite{Wilson2021quantumheuristics,Sung2020models} with inconclusive results.
More recently, the authors of~\cite{rudolph2021orqviz} introduced a method to visualize variational quantum landscapes through dimensionality reduction.
Yet landscape analysis tools from classical optimization have not been widely used to characterize quantum landscapes.

Landscape analysis aims at characterizing the landscape of cost functions by efficiently sampling the parameter space to understand the ``hardness'' of the optimization problem~\cite{10.1162/evco_a_00236,ZOU2022129,KerschkePWT15,MorganG17,BischlMTP12,kostovska2022per}.
For a VQA this implies only classical post-processing of data from a quantum device.
In contrast, the optimization step of a VQA involves constant interaction between quantum and classical resources.
In NISQ hardware such interactions might come with a large overhead.
To the best of our knowledge, no prior work on data-driven landscape analysis exists in the context of VQA.

In this work, we aim to close the gap between VQA and landscape analysis through the information content (IC)~\cite{munoz2015ELA} of the quantum landscape.
We demonstrate the connection between IC and the average norm of the gradient of the landscape.
We derive robust lower/upper bounds of this quantity which provides a crucial understanding of the landscape (e.g. complexity of optimizing the cost function).
We apply our results to numerically study the BP problem for local and global cost functions from ref.~\cite{cerezo2021cost}, showing excellent agreement with theoretical asymptotic scaling in the size of the gradient.
Also, we demonstrate how to calculate pre-factors of the asymptotic scaling, which are in practice more relevant for implementing algorithms.
As far as we know, this is the first work where scaling pre-factors are calculated in the context of VQAs and BPs.

The manuscript is organized as follows.
In~\Cref{sec:background} we give background on VQAs and IC.
We connect the average norm of the gradient with IC in~\Cref{sec:ic-gradients}, followed by a numerical diagnosis of BP using IC in ~\Cref{sec:bp_diagnose}.
\Cref{sec:prefactors} addresses the estimation of pre-factors in the scaling of BP.
In~\Cref{sec:conclusions} we discuss the implications of our results and point out future directions.

\section{Background}\label{sec:background}
\subsection{Parameterized Quantum Circuits}
In a variational quantum algorithm one aims at exploring the space of quantum states by minimizing a cost function with respect to a set of tunable real-valued parameters \(\vtheta \in [0, 2\pi)^m\) of a parametrized quantum circuit (PQC).
A PQC evolves an initial quantum state \(\initstate\) to generate a parametrized state
\begin{equation}\label{eq:state}
    \ket{\psi(\vtheta)} = U(\vtheta)\initstate,
\end{equation}
where \(U(\vtheta)\) is a unitary matrix of the form
\begin{equation}\label{eq:pqc}
U(\vtheta) = \prod_{i = 1}^m U_i (\vtheta_i) W_i, 
\end{equation}
with 
\begin{equation}
U_i (\vtheta_i) = \exp\left(-i \vtheta_i V_i\right).
\end{equation}
Here \(W_i\) are fixed unitary operations and \(V_i\) are hermitian matrices.
In a VQA these parameters are driven by (classically) minimizing a cost function \(C(\vtheta)\), built as the expectation value of a quantum observable \(\Obs\),
\begin{equation}\label{eq:cost}
C(\vtheta) = \bra{\psi_0} U^\dagger(\vtheta) \Obs U(\vtheta) \initstate.
\end{equation}
A successful optimization reaches an approximation to the lowest eigenvalue of \(\Obs\), and the optimal parameters represent an approximation to its ground-state~\cite{bharti2022noisy,cerezo2020variational}.
Our object of study is the manifold defined by a PQC and \(\Obs\) which we call a variational quantum landscape.

In order to measure \Cref{eq:cost} in a real device, one must prepare and measure the observable with multiple copies of \(U(\vtheta)\initstate\).
Therefore, the real cost function becomes
\begin{equation}
    \sampcost(\vtheta, R) = \langle \Obs \rangle + \kappa(R),
    \label{eq:sampled_cost_func}
\end{equation}
where \(\kappa(R)\) introduces the uncertainty of sampling the observable with \(R\) repetitions.
For sufficiently large number of them, \(\kappa(R)\) can be drawn from a Gaussian distribution with variance \(\sigma^2\sim 1/R\)~\cite{bonet-monroig2023performance}.
Throughout the rest of the text, we assume access to the exact value of the cost function, $C(\vtheta)$, unless otherwise stated.

\subsection{Exploratory Landscape Analysis} 
The goal of exploratory landscape analysis (ELA)~\cite{MersmannBTPWR11,KerschkeWPGDTE16,kerschke2019comprehensive} is to characterize a real-valued function  by numerically estimating a set of features from its landscape.
Examples of such features are multi-modality (the number of local minima), ruggedness, or curvature of the level set of a given landscape (see~\cite{kerschke2019comprehensive} for a comprehensive description of ELA features).
ELA becomes particularly useful when studying functions with unknown analytical form or exceedingly difficult for mathematical analysis where one can gain insight by comparing their features to those of known analytical functions~\cite{MorganG17,KerschkePWT15,MersmannPTBW15}.
Similarly, the classical machine learning/optimization community has made use ELA features to select the most suitable optimization algorithms to optimize unknown cost functions~\cite{KerschkeKBHT18,BischlMTP12,MersmannBT0BN13}.
A key aspect of landscape analysis is the fact that the number of evaluations of the cost function is significantly smaller than optimizing it; otherwise it is more efficient to optimize directly.

Among the ELA features, information content (IC)~\cite{munoz2015ELA,vassilev2000IC} is of particular interest as it characterizes the ruggedness of the landscape which is related to its trainability.
For example, high information content indicates a trainable landscape, whereas low information content indicates a flat landscape~\cite{kerschke2019comprehensive,munoz2015ELA}.
We demonstrate that for any classical or quantum variational landscape its trainability can be quantified with information content.

\subsection{Information content}\label{sec:ic}

\begin{definition}[Information Content (IC)]\label{def:ic}
Given a finite symbolic sequence $\phi = \{-, \odot, +\}^S$ of length $S$ and let \(p_{ab}, a \neq b \in \{-, \odot, +\}\) denote the probability that $ab$ occurs in the consecutive pairs of $\phi$. The information content is defined as
\begin{equation}\label{eq:ic}
    H = \sum_{a \neq b} h\left(p_{ab}\right),
    \end{equation}
    with
    \begin{equation}
        h(x) = -x \log_6 x.
    \end{equation}
\end{definition}
In this definition pairs of the same symbols are excluded, leaving only six combinations,
\begin{equation}\label{eq:p_six}
    p_{ab} =  \{p_{+-}, p_{-+}, p_{+\odot},p_{\odot +},p_{-\odot}, p_{\odot -}\}.
\end{equation}
The \(\log_6\) is necessary to ensure \(H\leq 1\).

To compute the IC, we use the algorithm given in ref.~\cite{munoz2015ELA}:
\begin{inparaenum}[(1)]
\item Sample \(M(m) \in \mathcal{O}(m)\) points of the parameter space \(\Theta = \{\vtheta_1, \dots, \vtheta_{M}\} \in [0, 2\pi)^m\).
\item Measure \(C(\vtheta_i)\) on a quantum computer (this is the only step where it is needed).
\item Generate a random walk \(W\) of \(S + 1 < M(m)\) steps over \(\Theta\), and compute the finite-size approximation of the gradient at each step \(i\)
\begin{equation}\label{eq:DeltaC}
    \Delta C_{i} = \frac{C(\vtheta_{i + 1}) - C(\vtheta_{i})}{\norm{\vtheta_{i + 1} - \vtheta_{i}}}.
\end{equation}
\item Create a sequence \(\phi(\epsilon)\) by mapping \(\Delta C_i\) onto a symbol in \(\{-, \odot, +\}\) with the rule
\begin{equation}\label{eq:discretization}
\phi(\epsilon) = \begin{cases}
- & \text{ if } \Delta C_i < -\epsilon \\
\odot & \text{ if } |\Delta C_i| \leq \epsilon \\
+ & \text{ if } \Delta C_i > \epsilon
\end{cases}
\end{equation}
\item Compute the empirical IC (denoted as \(H(\epsilon)\) henceforth) by applying~\Cref{def:ic} to \(\phi(\epsilon)\). 
\item Repeat these steps for several values of \(\epsilon\).
\end{inparaenum}

From this algorithm, we are interested in the regimes of high and low \(H(\epsilon)\)~\cite{vassilev2000IC,munoz2015ELA}.
The maximum IC (MIC) is defined as,
\begin{equation}\label{eq:mic}
    H_M = \max_{\epsilon} H(\epsilon),
\end{equation}
at \(\epsM\) as its corresponding \(\epsilon\).
The MIC occurs when the variability in the symbols of $\phi$ is maximum.

The other case of interest occurs when \(H(\epsilon) \leq \eta\) with $\eta \ll 1$.
This defines the sensitivity IC (SIC)~\cite{vassilev2000IC,munoz2015ELA},
\begin{equation}\label{eq:sic}
H_S=\min\left\{\epsilon>0 \vert  H(\epsilon) \leq \eta\right\}, 
\end{equation}
with $\epsilon=\epsS$.
The SIC identifies the $\epsilon$ at which (almost) all symbols in \(\phi\) are \(\odot\).
All symbols become exactly \(\odot\) at \(\eta = 0\).

Both MIC and SIC are calculated on a classical computer after collecting $(\vtheta, C(\vtheta))$.
The values $\epsilon_{M, S}$ can be found by sweeping over multiple values of \(\epsilon\) with logarithmic scaling, and computing \(H(\epsilon)\) with~\Cref{eq:ic} at each of those values\footnote{In the original work~\cite{kerschke2019comprehensive}, it is suggested to take $1000$ values of $\epsilon \in [0, 10^{15}]$}.
The cost is dominated by computing all quantities $\Delta C$ needed to construct $\phi(\epsilon)$ using~\Cref{eq:discretization}.
Since only $\bigO(m)$ samples are available, at most $\bigO(m^2)$ differences of the form $\Delta C$, can be computed. 
Thus, MIC and SIC can be obtained with at most $\bigO(m^2)$ classical operations. 

\section{Connection between information content and norm of the gradient}\label{sec:ic-gradients}
This section shows the relation between IC and the average norm of the gradient, from now denoted as \(\norm{\nabla C}\).
We take advantage of the fact that each step is isotropically random in \(W\).
This allows us to derive the underlying probability of \(\Delta C_i\).
Additionally, we use IC to bound the probability of pairs of symbols appearing along \(W\), which allows us to estimate \(\norm{\nabla C}\). 
Although we demonstrate our results for a variational quantum landscape, they extend to any optimization landscape.

\subsection{Estimation of the norm of the gradient}\label{sec:gradients}
The random walks \(W\) over \(\Theta\) satisfy
\begin{align}
    \norm{\vtheta_{i+1} - \vtheta_{i}} & \leq d \\
    \frac{\vtheta_{i+1} - \vtheta_{i}}{\norm{\vtheta_{i+1} - \vtheta_{i}}} & = \vec{\delta}_{i}, 
\end{align}
where $\vec\delta_i$ is drawn from the isotropic distribution and \(d\) is fixed before starting the walk but might be varied.
By Taylor expanding~\Cref{eq:DeltaC} and the mean-value theorem, the finite-size gradient can be written as 
\begin{equation}\label{eq:mean-value}
\Delta C_i = \nabla C((1 - t)\vtheta_i + t \vec\theta_{i+1} )\cdot \vec\delta_i,
\end{equation}
with $t \in (0, 1)$.
Since the sampled points \(\Theta\) are chosen randomly, we can assume that \(\Delta C\) and \( \nabla C(\vtheta) \cdot \vec\delta \) are drawn from the same probability distribution, given a sufficiently large \(\Theta\).

The isotropic condition of \(W\) allows us to calculate the probability distribution of \(\nabla C(\vtheta) \cdot \vec\delta\):
\begin{lemma}\label{le:beta}name
    Let $C(\vtheta)$ be a differentiable function for all $\vtheta\in [0, 2\pi)^m$.
    Let $\vec\delta \in \mathbb{R}, \Vert \vec\delta\Vert = 1$ be drawn from the isotropic distribution.
    Then $(\nabla C( \vtheta)\cdot\vec{\delta})^2$ is a random variable with a beta  probability distribution $(\Beta)$~\cite{bailey1992distributional} such that
\begin{equation}
    \left(\nabla C( \vtheta)\cdot\vec{\delta}\right)^2\sim \norm{\nabla C(\vtheta)}^2\Beta\left(\frac{1}{2}, \frac{m-1}{2}\right). 
\end{equation}
\end{lemma}
The proof can be found in~\Cref{app:beta}. 

We can use \Cref{le:beta} to bound the probability of \(\nabla C(\vtheta)\) from the \(\Beta\) cumulative distribution function (CDF).
\begin{theorem}[CDF of \(\nabla C( \vtheta)\cdot\vec{\delta}\)]\label{th:cdf}
 Let $C(\vtheta)$ be a differentiable function at every $\vtheta\in [0, 2\pi)^m$.
 Let $\vec\delta \in \mathbb{R}, \Vert \vec\delta \Vert = 1$ be drawn from the isotropic distribution.
 Then $\nabla C( \vtheta)\cdot\vec{\delta}$ is a random variable with cumulative density function 
    \begin{multline}
    \operatorname{\rm Prob}\left(\nabla C( \vtheta)\cdot\vec{\delta} \leq \epsilon \right) = \\
        \frac{1}{2}\left(1 + \operatorname{sgn}(\epsilon) \ \I \left(\frac{\epsilon^2}{\norm{C(\vtheta)}^2};\frac{1}{2}, \frac{m-1}{2}\right)\right),
\end{multline}
where $\I(x;\alpha, \beta)$ is the regularized incomplete beta function with parameters $\alpha$ and $\beta$.
\end{theorem}
The proof of this theorem can be found in~\Cref{app:cdf}.

It is known that the beta distribution (with the parametrization in~\Cref{le:beta}) rapidly converges to a normal distribution~\cite{temme1992asymptotic}:
\begin{equation}\label{eq:gaussian}
\lim_{m\rightarrow\infty} \nabla C( \vtheta)\cdot\vec{\delta} \sim \mathcal{N}\left(0, \frac{\norm{C(\vtheta)}^2}{m}\right).
\end{equation}
This approximation is accurate even for reasonably small values of $m$.
We can naturally interpret the functionality of $\sqrt{m}$ as dimensionality normalization in~\Cref{le:beta}.

\Cref{le:beta} implies, in the Gaussian limit,
\begin{equation}
    \E_{W}\left(C(\vtheta)\cdot\vec\delta\right) \sim 
    \mathcal{N}\left(0, \frac{\Normgrad_W^2}{m} \right)
\end{equation}
where $\E_W$ denotes the expectation taken over the points in random walk $W$, and 
\begin{equation}\label{eq:average_norm}
    \Normgrad^2_W = \E_W\left(\norm{\nabla C(\vtheta)}^2\right).
\end{equation}
As an immediate consequence, we give the CDF of $\nabla C(\vtheta)\cdot\vec\delta$ averaged over a random walk $W$:
\begin{corollary}[CDF of average norm of gradients]\label{cor:cdf}
 Let $C(\vtheta)$ be a differentiable function at every $\vtheta\in [0, 2\pi)^m$.
 Let $\vec\delta \in \mathbb{R}, \Vert \vec\delta \Vert = 1$ be drawn from the isotropic distribution.
 Then $\nabla C( \vtheta)\cdot\vec{\delta}$ is a random variable with cumulative density function 
\begin{equation}
        \operatorname{\rm Prob}\left(\E_W\left(\nabla C(\vtheta)\cdot\vec{\delta}\right) \leq \epsilon \right) = \operatorname{\Phi}_G\left(\frac{\epsilon\sqrt{m}}{\Normgrad}\right),
\end{equation}
where $\Phi_G(\cdot)$ is the CDF of the standard normal distribution.
\end{corollary}

Note that $\Normgrad_W$ converges to the average norm of the gradient over $[0, 2\pi)^m$, i.e., $\Normgrad^2\coloneqq\E||\nabla C(\vtheta)||^2$ at a rate~\cite{hastings1970monte}:
\begin{equation}\label{eq:normgrad-converge}
    \bigg\vert\Normgrad - \Normgrad_W  \bigg\vert\in\mathcal{O}_p(M^{-1/2}),
\end{equation}
allowing us to approximate the interesting $\Normgrad$ with the accessible $\Normgrad_W$ with small error. 

\subsection{Probability concentration for information content}\label{sec:ic_prob}
Our next goal is to bound the probability of pairs of symbols appearing in \(\phi\) in the high and low IC regimes.

\subsubsection*{High information content}

If one interprets IC as a partial entropy of the landscape, high \(H\) necessarily implies approximately equal probabilities \(p_{ab}\).
Therefore, a minimal concentration of probabilities must exist such that a high value of \(H\) can be reached.
One can formally define this statement;
\begin{lemma}\label{le:h_max}
    Let $H > 2 h(1/2)$ be the IC of a given sequence $\phi$.
    Consider the probabilities in~\Cref{eq:p_six} such that the sum of any four of them (\(p_4\)) is bounded by
    \begin{equation}
    p_4 \geq 4 q,
    \end{equation}
    with $q$ the solution of $H = 4 h(x) + 2 h(1/2 - 2x)$. 
\end{lemma}
The proof can be found in~\Cref{app:h_max}. The bound on \(p_4\) in the above lemma gets tighter as \(q\) increases, and so does \(H\), which by definition \(H \leq 1\).
The tightest possible bound is achieved for the MIC defined in~\Cref{eq:mic}.

\subsubsection*{Low information content}
For a low IC to occur, all \(p_{ab}\) must be small, and their values can be upper-bounded.
\begin{lemma}\label{le:h_s}
    Let $H$ be the IC with bound $H \leq \eta \leq 1 / 6$ of a given sequence $\phi$.
    Then the probability of consecutive \(\odot\) steps during a random walk are close to $1$.
    The expected norm of the gradient is bounded by
    \begin{equation}
    p_{\odot \odot} \geq 1 - 3 \eta, 
\end{equation}
\end{lemma}
The proof can be found in~\Cref{app:h_s}. As in the previous case, a tight bound is attained with the SIC defined in~\Cref{eq:sic}.

High and low IC provides insight into the hardness to optimize the cost function that generated the landscape.
As shown in this section, low IC implies a flat landscape, which clearly imposes hardness in optimization.
In contrast, high IC is a necessary but not sufficient condition for optimization.
In this case there is a guarantee that the landscape can be, at the very least, easy to optimize locally.
However, IC does not provide any information on the quality of the accessible minimum (e.g. global or local) or the multi-modality (number of local minima) of the landscape. The reason is the random walk over $\Theta$, which allows for high IC both for convex or multimodal landscapes. 

\subsection{Information content to estimate the norm of the gradient}
We are ready to show the main result of this work.
We make use of the results in~\Cref{sec:gradients} and~\Cref{sec:ic_prob} to prove that $H(\epsilon)$ estimates the average norm of the gradient $\Normgrad$ for any classical or quantum landscape.
To the best of our knowledge, this is the first time, including the field of classical optimization, where such bounds are calculated.

First, we relate \(H_M\) to the norm of the gradient.
High values of IC guarantee a minimal probability for individual steps to increase and decrease and thus bounds the compatible values of $\epsilon / \Normgrad_W$.
\begin{theorem}[$H_M$ bounds $\Normgrad_W$]\label{th:h_m_grad}
Let $H_M$ be the empirical MIC of a given function $C(\vtheta)$, and $\epsM$ its corresponding $\epsilon$.
Let $q$ be the solution to the equation $H_M = 4 h(x) + 2 h(1/2 - 2x)$.
Then,
\begin{equation}\label{eq:bound_epsmax}
     \frac{\epsM \sqrt m}{\Phi_{G}^{-1}(1 - 2 q)} \leq \Normgrad_W\leq \frac{\epsM \sqrt m}{\Phi_{G}^{-1}\left( \frac{1 + 2q}{2}\right)}.
\end{equation}
\end{theorem}
The proof is given in~\Cref{app:h_m_grad}.
 
The second result connects SIC to the bounds in the norm of the gradient.
Small values of IC imply a large probability of consecutive \(\odot\) steps in \(\phi_i\) or equivalently small probabilities for \(p_{ab}\).
When this occurs, then \(\epsS\) is used to upper bound \(\Normgrad_W\).
\begin{theorem}[$H_S$ upper bounds $\Normgrad_W$]\label{th:h_s_grad}
Let $H_S$ be the empirical SIC of a given function $C(\vtheta)$, and $\epsS$ its corresponding $\epsilon$.
Then 
\begin{equation}\label{eq:bound_eps_s}
\Normgrad_W \leq \frac{\epsS \sqrt m}{\Phi_{G}^{-1}\left(1-3\eta / 2\right)}.
\end{equation}
\end{theorem}
The proof is given in~\Cref{app:h_s_grad}.

From~\Cref{eq:normgrad-converge}, it is obvious that $\Normgrad_W$ well approximate $\Normgrad$ for a large $M$. Hence, we can use~\Cref{th:h_m_grad} and~\Cref{th:h_s_grad} to bound $\Normgrad$ with a long sequence $\phi$.  

\Cref{th:h_m_grad} and~\Cref{th:h_s_grad} provide confidence intervals of \(\Normgrad\) without prior knowledge of the variational landscape. The former gives both an upper and lower bound of \(\Normgrad\) but can only be applied when \(H\) has a minimum value, while the latter is always applicable with an arbitrary small \(\eta\) for an upper bound.

Finally, we discuss the tightness of the bounds in~\Cref{th:h_m_grad} and~\Cref{th:h_s_grad}.
Firstly, the bound in~\Cref{cor:cdf} might become loose if the approximation from the Beta to a Gaussian distribution is not accurate.
However, the error between these distributions is negligible already at \(m \sim 10\).
Moreover, as the value of \(m\) increases, so does the distribution, and the bound gets tighter.
Another possibility might come from the entropy arguments of IC (see~\Cref{le:h_max} and~\Cref{le:h_s}).
The bound saturates at $H_M = 1$ but becomes looser as $H_M$ decreases.
When $H\leq 2h(1/2)$, they are no longer trustworthy.
Nonetheless, for common values of $H_M\sim 0.8$, the bounds are only loose by small factors.
Similarly for $H_S$, the bound tightens as $\eta$ decreases but becomes uncontrolled with \(\eta > 1/6\). 

\subsection{Information content under sampling noise}
In the presence of sampling noise, symbols in \(\phi(\epsilon)\) might be flipped due to the uncertainty of the cost function.
The amount of flips within the sequence depends on the number of repetitions \(R\). The following condition ensures that symbols in the sequence will not be flipped with high probability:
\begin{equation}
    R\geq \epsilon^{-2} \norm{\vtheta_{i + 1} - \vtheta_{i}}^{-2} \geq \epsilon^{-2}d^{-2}.
\end{equation}
It induces a lower bound $\epsilon \geq 1 / d\sqrt{R}$. 
In this scenario, the bound in~\Cref{cor:cdf} and all subsequent results transform
\begin{equation}
    \epsilon \rightarrow \epsilon \pm \kappa(R).
\end{equation}
The bounds become slightly looser from the sampling noise dependence.

Sampling noise posses the same limitation to both IC and optimization.
If $\Normgrad$ decays exponentially, then exponentially many repetitions are required to resolve the gradient.
Yet, IC is capable of signaling non-resolvable (flat) landscapes with less evaluations than a full optimization.

\section{Information content to diagnose barren plateaus}\label{sec:bp_diagnose}
Our goal is now to apply the previous results to study the problem of barren plateaus (BP)~\cite{mcclean2018barren,cerezo2021cost}
We choose this problem because there exist analytical results on the scaling of the \(\Normgrad\).
This allows us to directly verify that \(H(\epsilon)\) can be used as a proxy to \(\Normgrad\).

BPs are characterized by the following conditions~\cite{mcclean2018barren},
\begin{align}
\mathbb E \left( \partial_k C(\vtheta)\right) & = 0, \\
\Var \left( \partial_k C(\vtheta)\right) & \in \mathcal O \left(\exp(-n) \right),
\end{align}
where \(n\) is the number of qubits.
BP implies exponentially vanishing variances of the derivatives.
Similarly, BP can be understood as having a flat optimization landscape.
These two concepts are connected by Chebyshev's inequality~\cite{tchebychef1867valeurs}.

The IC allows us to calculate
\begin{equation}
    \Normgrad \approx \E_W\left(\sum_{k = 1}^m(\partial_k C(\vtheta))^2\right) = \sum_{k = 1}^m \Var_W(\partial_k C(\vtheta)),
\end{equation}
where $\Var_W$ computes the variance over points of the random walk $W$.
Hence, IC is a proxy of the average variance of each partial derivative in the parameter space.

\section{Numerical experiments}\label{sec:numerics}
To showcase IC as a tool to analyze the landscape of a VQA, we perform a numerical study of the BP problem as described by Cerezo et al. in~\cite{cerezo2021cost}.
Here, the authors analytically derive the scaling of \(\Var(\partial C(\vec\theta))\) in two different scenarios. Such scaling depends both on the qubit size and circuit depth of the PQC. If the cost function is computed from global observables (e.g. non-trivial support on all qubits), BPs exist irrespective of the depth of the PQC. In the case of local observables (e.g. non-trivial support on a few qubits), one can train shallow PQCs, but BPs gradually appear as the circuit depth increases.
These results hold for alternating layered ansatze composed of blocks of 2-local operations (Fig.~4 in~\cite{cerezo2021cost}).

In our numerical experiments, we use circuits from 2 to 14 qubits, each of them going from 4 to 16 layers.
We calculate the cost function from
\begin{align}\label{eq:observables}
    \Olocal =\frac{1}{n} \sum_{i = 1}^n (Z_i - 1)\\
    \Oglobal =\bigotimes_{i = 1}^n \ket 0 \bra 0^{\otimes n},
\end{align}
without sampling noise.
Further details of the numerical experiments are given in~\Cref{app:numerical_experiments}.

\begin{figure}
    \centering
    \includegraphics[width=\columnwidth]{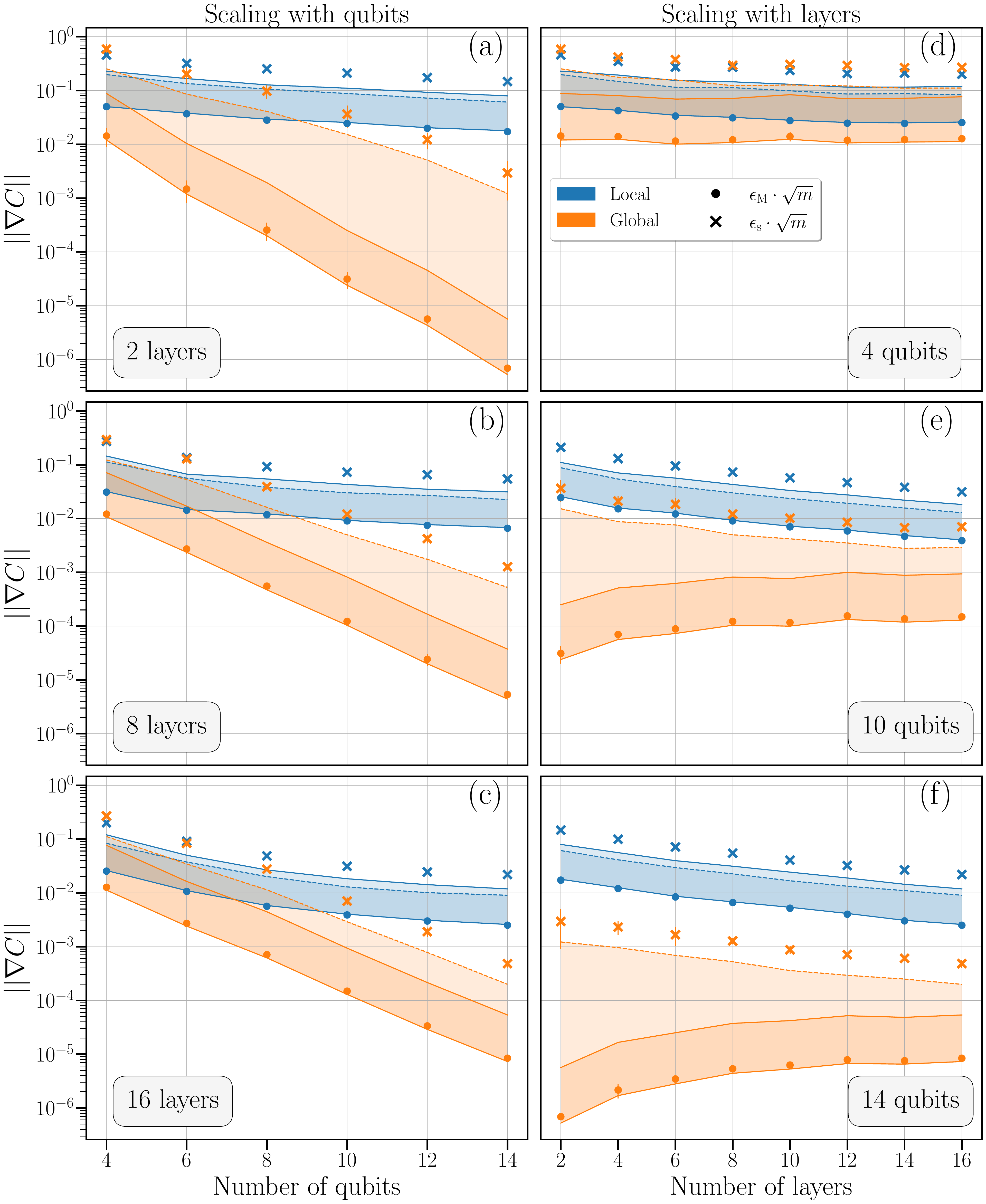}
    \caption{Scaling of the average gradient \(\Normgrad\).
    Panels (a,b,c) shows the scaling with with respect to qubits, and panels (d,f,g) the scaling with respect to layers.
    The solid lines show the bounds from~\Cref{eq:bound_epsmax} (solid lines) and~\Cref{eq:bound_eps_s} (dashed lines).
    \(\Normgrad\) can take values within the shadow areas in between these lines.
    The markers refer to the values of \(\epsMp\) (dots) and \(\epsS \cdot \sqrt{m}\) (crosses).
    They are calculated from the median of five independent runs, with their standard deviation as the error bars.
    The colors represent the results for the local (blue) and global (orange) cost functions.
    }\label{fig:scaling_qubits_layers}

    \centering
    \includegraphics[width=\columnwidth]{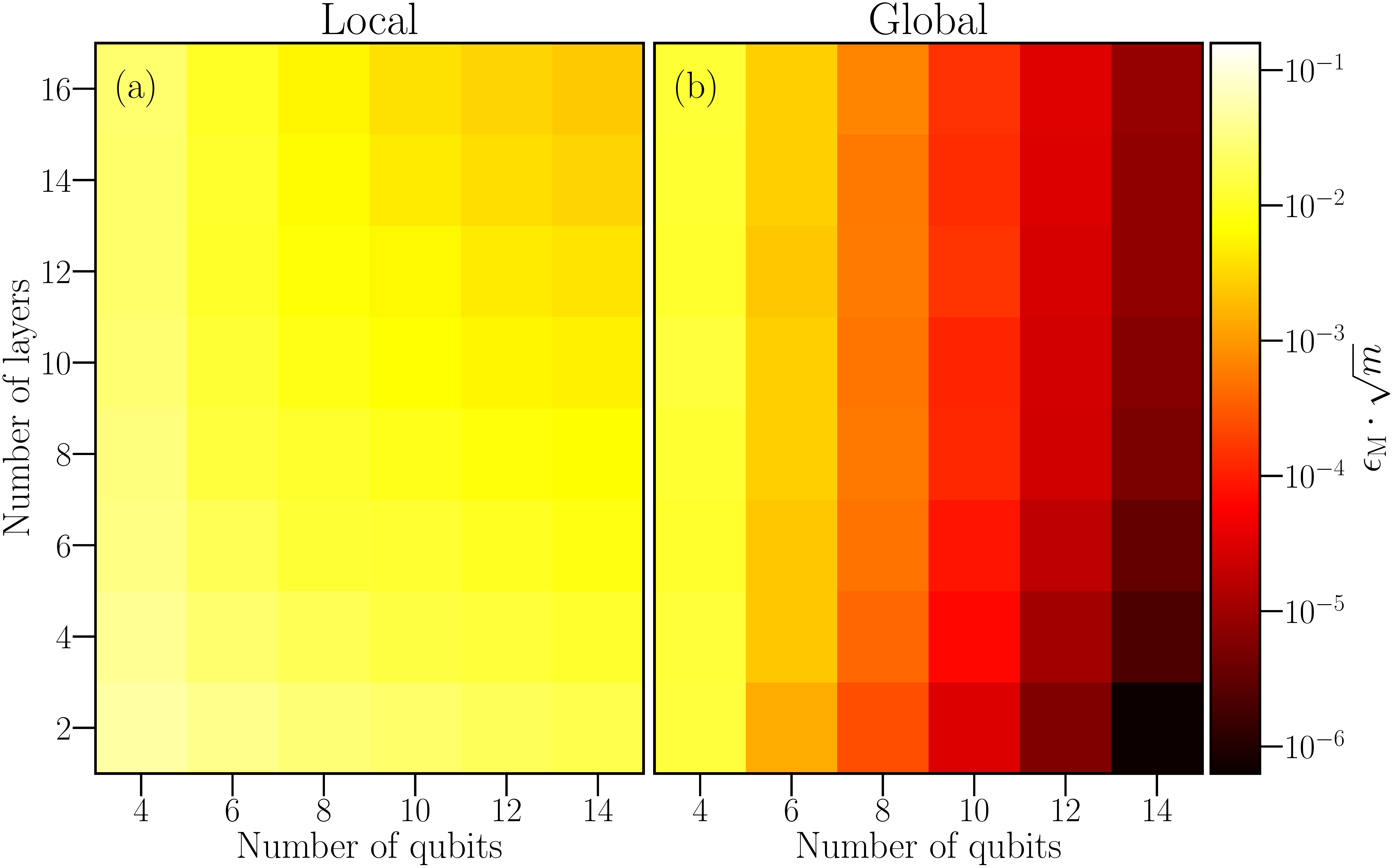}
    \caption{Heatmap of \(\epsMp\).
    Panel (a) shows the values for the local cost function, and panel (b) the global cost function.
    The number of qubits is depicted on the x-axis, while the number of layers is shown on the y-axis.}
    \label{fig:local_vs_global}
\end{figure}

The results of the BP problem using IC are shown in~\Cref{fig:scaling_qubits_layers}.
In all plots we compute the bounds on the \(\Normgrad\) from~\Cref{th:h_m_grad} (solid lines) and~\Cref{th:h_s_grad} (dashed lines), for the local (blue) and global (orange) cost functions.
Additionally we show the value of \(\epsMp\) (dots) and \(\epsS \cdot \sqrt{m}\) (crosses).

The first trend we observe is that the \(\Normgrad\) shows two different scalings with respect to qubits (left panels) and layers (right panels).
The scaling with qubits shows a \(\mathcal{O}({\rm poly^{-1}}(n))\) decay in the local cost function and a remarkable \(\mathcal{O}({\exp(-n)})\) decay with the global cost function.
We emphasize the fact that these results are in perfect agreement with the predictions in~\cite{cerezo2021cost}.
On the other hand, the scaling with layers strongly depends on the number of qubits.
For 4 qubits, \(\Oglobal\) has a constant value \(\Normgrad \approx 10^{-2}\), while \(\Olocal\) shows a small decay with a similar average.
For 10 and 14 qubits, we recover the predicted \(\mathcal{O}({\rm poly^{-1}}(n))\) decay in the \(\Olocal\).
In contrast, \(\Oglobal\) has \(\Normgrad \approx 10^{-4} - 10^{-6}\) \(\Oglobal\) which is already close to float precision.
Finally, we observe that \(\epsMp\) is close to the lower bound, thus making it a robust proxy for \(\Normgrad\).

In~\Cref{fig:local_vs_global} we show a heatmap of the values of \(\epsMp\) when increasing the number of qubits (x-axis) and the layers (y-axis) for both local (left) and global (right) cost function.
The values of results for \(\Olocal\) (left panel) show a rich variety of features: for 2 to 6 layers \(\epsMp\) shows a very mild decay, but for more than 8 layers the decay sharpens.
This is exactly as expected for local cost functions: BPs appear gradually as the circuit depth increases.
We speculate that the color change at the top right corner of the left panel in~\Cref{fig:local_vs_global} corresponds to a transition regime.
With regard to the global cost function (panel b), the expected exponential decay (in the number of qubits) is observed.

Surprisingly, both~\Cref{fig:scaling_qubits_layers} and~\ref{fig:local_vs_global} show an increase in \(\epsMp\) (or equivalently \(\Normgrad\)) at a fixed number of qubits as the number of layers grows for the global cost function.
We have not been able to find an explanation for this behavior either analytically or in the literature.
However, this is an example of how data-driven methods might provide useful insight for deeper understanding.

\section{Estimation of scaling pre-factor}\label{sec:prefactors}

\begin{table}[h!]
\begin{tabular}{|ccccc|}
\hline
\multicolumn{5}{|c|}{Global cost function prefactors}                                                                       \\ 
\multicolumn{5}{|c|}{\(f(n) = 2^{\alpha n + \beta}\)}                                                                       \\\hline
\multicolumn{1}{|c|}{}   & \multicolumn{2}{c|}{\(\alpha\)}                              & \multicolumn{2}{c|}{\(\beta\)}          \\ \hline
\multicolumn{1}{|c|}{Layers} & \multicolumn{1}{c|}{LB}    & \multicolumn{1}{c|}{\(\epsMp\)}  & \multicolumn{1}{c|}{LB}    & \(\epsMp\)  \\ \hline
\multicolumn{1}{|c|}{2}      & \multicolumn{1}{c|}{-1.43} & \multicolumn{1}{c|}{-1.41} & \multicolumn{1}{c|}{-0.91} & -0.68 \\ \hline
\multicolumn{1}{|c|}{4}      & \multicolumn{1}{c|}{-1.29} & \multicolumn{1}{c|}{-1.27} & \multicolumn{1}{c|}{-1.22} & -1.09 \\ \hline
\multicolumn{1}{|c|}{6}      & \multicolumn{1}{c|}{-1.19} & \multicolumn{1}{c|}{-1.17} & \multicolumn{1}{c|}{-1.85} & -1.68 \\ \hline
\multicolumn{1}{|c|}{8}      & \multicolumn{1}{c|}{-1.13} & \multicolumn{1}{c|}{-1.12} & \multicolumn{1}{c|}{-1.98} & -1.85 \\ \hline
\multicolumn{1}{|c|}{10}     & \multicolumn{1}{c|}{-1.12} & \multicolumn{1}{c|}{-1.12} & \multicolumn{1}{c|}{-1.97} & -1.82 \\ \hline
\multicolumn{1}{|c|}{12}     & \multicolumn{1}{c|}{-1.06} & \multicolumn{1}{c|}{-1.05} & \multicolumn{1}{c|}{-2.41} & -2.26 \\ \hline
\multicolumn{1}{|c|}{14}     & \multicolumn{1}{c|}{-1.07} & \multicolumn{1}{c|}{-1.07} & \multicolumn{1}{c|}{-2.29} & -2.14 \\ \hline
\multicolumn{1}{|c|}{16}     & \multicolumn{1}{c|}{-1.06} & \multicolumn{1}{c|}{-1.06} & \multicolumn{1}{c|}{-2.25} & -2.10 \\ \hline
\end{tabular}
\caption{Estimated qubit scaling pre-factors of the global cost function by linear fitting. 
\(\alpha\) is the slope and \(\beta\) the intercept of the linear model.
Each coefficient has two columns showing the results of fitting \(\epsMp\) and its lower bound (LB).
}
\label{tab:global_scaling_qubits}
\vskip2mm
\begin{tabular}{|ccccccc|}
\hline
\multicolumn{7}{|c|}{Local cost function scaling with qubits}                                                                                                                                   \\ 
\multicolumn{7}{|c|}{\(f^{-1}(n) = \alpha n^{2} + \beta n + \gamma\)}                                                                                                                                   \\ \hline
\multicolumn{1}{|c|}{}   & \multicolumn{2}{c|}{\(\alpha\)}                               & \multicolumn{2}{c|}{\(\beta\)}                               & \multicolumn{2}{c|}{\(\gamma\)}             \\ \hline
\multicolumn{1}{|c|}{Layers} & \multicolumn{1}{c|}{LB}    & \multicolumn{1}{c|}{\(\epsMp\)}  & \multicolumn{1}{c|}{LB}    & \multicolumn{1}{c|}{\(\epsMp\)}  & \multicolumn{1}{c|}{LB}      & \(\epsMp\)    \\ \hline
\multicolumn{1}{|c|}{2}      & \multicolumn{1}{c|}{0.05}  & \multicolumn{1}{c|}{0.03}  & \multicolumn{1}{c|}{2.80}  & \multicolumn{1}{c|}{3.3}   & \multicolumn{1}{c|}{8.05}    & 6.37    \\ \hline
\multicolumn{1}{|c|}{4}      & \multicolumn{1}{c|}{-0.25} & \multicolumn{1}{c|}{-0.24} & \multicolumn{1}{c|}{10.08} & \multicolumn{1}{c|}{10.22} & \multicolumn{1}{c|}{-12.86}  & -12.96  \\ \hline
\multicolumn{1}{|c|}{6}      & \multicolumn{1}{c|}{-0.16} & \multicolumn{1}{c|}{-0.16} & \multicolumn{1}{c|}{11.09} & \multicolumn{1}{c|}{11.52} & \multicolumn{1}{c|}{-11.37}  & -12.16  \\ \hline
\multicolumn{1}{|c|}{8}      & \multicolumn{1}{c|}{-0.27} & \multicolumn{1}{c|}{-0.30} & \multicolumn{1}{c|}{16.20} & \multicolumn{1}{c|}{16.99} & \multicolumn{1}{c|}{-26.22}  & -28.13  \\ \hline
\multicolumn{1}{|c|}{10}     & \multicolumn{1}{c|}{-0.56} & \multicolumn{1}{c|}{-0.55} & \multicolumn{1}{c|}{25.33} & \multicolumn{1}{c|}{25.63} & \multicolumn{1}{c|}{-57.30}  & -57.50  \\ \hline
\multicolumn{1}{|c|}{12}     & \multicolumn{1}{c|}{-0.31} & \multicolumn{1}{c|}{-0.34} & \multicolumn{1}{c|}{25.52} & \multicolumn{1}{c|}{26.77} & \multicolumn{1}{c|}{-56.33}  & -59.65  \\ \hline
\multicolumn{1}{|c|}{14}     & \multicolumn{1}{c|}{0.14}  & \multicolumn{1}{c|}{0.08}  & \multicolumn{1}{c|}{25.99} & \multicolumn{1}{c|}{28.00} & \multicolumn{1}{c|}{-69.42}  & -75.87  \\ \hline
\multicolumn{1}{|c|}{16}     & \multicolumn{1}{c|}{0.13}  & \multicolumn{1}{c|}{0.12}  & \multicolumn{1}{c|}{33.69} & \multicolumn{1}{c|}{34.95} & \multicolumn{1}{c|}{-104.77} & -108.92 \\ \hline
\end{tabular}
\caption{Estimated qubit scaling pre-factors of the local cost function by fitting a second-degree polynomial.
Each coefficient has two columns showing the results of fitting \(\epsMp\) and its lower bound (LB).
}
\label{tab:local_scaling_qubits}
\vskip2mm
\begin{tabular}{|ccccccc|}
\hline
\multicolumn{7}{|c|}{Local cost function scaling with layers}                                                                                                                                                 \\
\multicolumn{7}{|c|}{\(f^{-1}(l) = \alpha l^{2} + \beta l + \gamma\)}                                                                                                                                                 \\\hline
\multicolumn{1}{|c|}{}   & \multicolumn{2}{c|}{\(\alpha\)}                               & \multicolumn{2}{c|}{\(\beta\)}                             & \multicolumn{2}{c|}{\(\gamma\)}                             \\ \hline
\multicolumn{1}{|c|}{Qubits} & \multicolumn{1}{c|}{LB}    & \multicolumn{1}{c|}{\(\epsMp\)}  & \multicolumn{1}{c|}{LB}   & \multicolumn{1}{c|}{\(\epsMp\)} & \multicolumn{1}{c|}{LB}    & \multicolumn{1}{c|}{\(\epsMp\)} \\ \hline
\multicolumn{1}{|c|}{4}      & \multicolumn{1}{c|}{-0.1}  & \multicolumn{1}{c|}{-0.11} & \multicolumn{1}{c|}{3.29} & \multicolumn{1}{c|}{3.5}  & \multicolumn{1}{c|}{12.74} & 12.43                     \\ \hline
\multicolumn{1}{|c|}{6}      & \multicolumn{1}{c|}{-0.27} & \multicolumn{1}{c|}{-0.28} & \multicolumn{1}{c|}{9.54} & \multicolumn{1}{c|}{9.94} & \multicolumn{1}{c|}{6.26}  & 6.10                      \\ \hline
\multicolumn{1}{|c|}{8}      & \multicolumn{1}{c|}{0.13}  & \multicolumn{1}{c|}{0.13}  & \multicolumn{1}{c|}{7.69} & \multicolumn{1}{c|}{7.98} & \multicolumn{1}{c|}{18.80} & 19.16                     \\ \hline
\multicolumn{1}{|c|}{10}     & \multicolumn{1}{c|}{0.48}  & \multicolumn{1}{c|}{0.52}  & \multicolumn{1}{c|}{5.98} & \multicolumn{1}{c|}{5.89} & \multicolumn{1}{c|}{27.79} & 29.72                     \\ \hline
\multicolumn{1}{|c|}{12}     & \multicolumn{1}{c|}{0.81}  & \multicolumn{1}{c|}{0.82}  & \multicolumn{1}{c|}{4.68} & \multicolumn{1}{c|}{5.04} & \multicolumn{1}{c|}{38.24} & 38.70                     \\ \hline
\multicolumn{1}{|c|}{14}     & \multicolumn{1}{c|}{1.14}  & \multicolumn{1}{c|}{1.18}  & \multicolumn{1}{c|}{2.95} & \multicolumn{1}{c|}{3.04} & \multicolumn{1}{c|}{48.93} & 50.89                     \\ \hline
\end{tabular}
\caption{Estimated layer scaling pre-factors of the local cost function by fitting a second-degree polynomial.
Each coefficient has two columns showing the results of fitting \(\epsMp\) and its lower bound (LB).
}
\label{tab:local_scaling_layers}
\end{table}

Thus far the numerical results have just confirmed the asymptotic theoretical predictions of the considered BP problem.
Our methodology can be used beyond the asymptotic scaling to compute actual pre-factors by fitting \(\epsMp\) to its predicted functional form, including bounds on them from~\Cref{eq:bound_epsmax}.
Obtaining such pre-factors is challenging analytically.
Yet they are relevant when studying the complexity of an algorithm in practice.
In this section, we obtain the scaling pre-factors for the global cost function (in the number of qubits) and the local cost function for the number of qubits and layers (see \Cref{app:numerical_experiments} for additional details of these fittings).

First, we study the global cost function scaling with qubits for each number of layers in our data.
We fit a linear model \(f(x) = \alpha x + \beta\) with \(x = \rm{log}_2(\epsMp)\).
The results of the fit are shown in~\Cref{tab:global_scaling_qubits}.
For each of the coefficients, we show the fitting values of the lower bound (LB column) and \(\epsMp\) (right column).
As the number of layers increases, \(\alpha \rightarrow -1.0\), which is consistent with exponential decay of the form \(2^{-n}\) predicted in~\cite{cerezo2021cost}.
More importantly, asymptotic scaling is not sensitive to the constant factor \(\beta\), but it is given by the right column in~\Cref{tab:global_scaling_qubits}. 
Based on the trend in this column we speculate that the constant factor is \(\beta\rightarrow -2.0\).

In the case of \(\Olocal\), there are fewer known asymptotic predictions of the gradient norm.
In~\cite{cerezo2021cost} it is shown that there exist three regimes: trainable, BPs, and transition area, depending on the depth with respect to the system size.
Due to the small number of qubits and layers of our numerical study, we assume to be in the trainable regime, where theory predicts a \(\Normgrad\) scaling in \(\mathcal{O}({\rm{poly}^{-1}}(n))\).
We use a second-order polynomial model \(f(n) = \alpha n^{2} + \beta n + \gamma\) to fit \((\epsMp)^{-1}\).
The results are given in \Cref{tab:local_scaling_qubits}.
The first observation is the small value of the quadratic coefficient for all layers.
This might lead to thinking that a linear function will be better suited.
To discard this possibility we perform a linear fit (see \Cref{fig:app_local_qubit_scaling_fit} in \Cref{app:numerical_experiments}) leading to comparable values of the slope and intercept, with slightly better fitting statistics for the second-degree polynomial.
The \(\beta\) coefficient shows a 10-fold increase as the number of layers grows.
In contrast, \(\gamma\) gets increasingly more negative with the number of layers.
Note that we can extract the degree of the polynomial, which is impossible from the theory.

Lastly, we estimate the scaling coefficients of the local cost function with respect to layers, where no theoretical scaling is known~\cite{cerezo2021cost}.
We choose a second-order polynomial as the hypothesis functional form to fit the data.
A linear model clearly under-fits the data (see~\Cref{fig:app_local_layer_scaling_fit}), thus confirming the intuition of a higher-order polynomial scaling.
The results are presented in \Cref{tab:local_scaling_layers}.
The quadratic coefficient \(\alpha\) increases as the number of qubits grows at \(n \geq 8\), and so does the constant term \(\gamma\).
In contrast, the linear term \(\beta\) remains roughly constant across all system sizes studied.
An opposite trend in the coefficients seems to occur between \(n=4\) and \(n=6\), \(\alpha\) and \(\gamma\) decrease, while \(\beta\) increases.
We have not been able to match this change in the tendency to a change in scaling, leaving a finite-size effect in the fitting as the most possible explanation.

The results presented in this section are a demonstration that data-driven approaches can provide useful insight to complement analytical methods and can be leveraged to get a deeper understanding of a problem. 

\section{Conclusions}\label{sec:conclusions}
Variational quantum algorithms have been intensively studied as a suitable application for near-term quantum hardware.
From a computer science perspective, they are simply an optimization problem with classical input/output, yet their cost function is a quantum object.
The parameters of any optimization problem induce a landscape that contains information about its ``hardness''.
Landscape analysis is central to classical optimization but has been somewhat ignored in the VQAs community.

In this work, we investigate the information content features of a variational quantum landscape, which can be calculated efficiently by sampling the parameter space.
We prove that for any (classical or quantum) cost function the average norm of its gradient can be rigorously bounded with the information content. 
We validate our theoretical understating by a numerical experiment, confirming the predicted asymptotic scaling of the gradient in the barren plateau problem.
Finally, we apply our results to predict scaling pre-factors of the gradient in a data-driven fashion.
To our knowledge, this is the first time that such pre-factors are calculated for a VQA.

The study of optimization landscapes of VQA opens a new avenue to explore their capabilities within the NISQ era.
First, landscape analysis does not require constant interaction between quantum and classical hardware.
Secondly, only linear (in the number of parameters) queries to a quantum computer suffice to extract the information content instead of polynomially many queries for a standard optimization routine.
Finally, information content might be used as an easy and comparable metric between ansatzes.

We envision future research directions with information content such as studying the feasibility of the VQA optimization, estimating the number of shots needed to resolve a gradient, or warm-starting the algorithm from regions of interest in parameter space. 
Importantly, landscape analysis does not rely on any constraint in the quantum circuit.
It can be then used even in the case when analytical approaches are unavailable.
Therefore we anticipate that landscape analysis and information content might have a broad range of applications beyond VQAs in the NISQ era. 

\section*{Data availability}
The data to reproduce the results of this work can be found in \href{https://doi.org/10.5281/zenodo.7760281}{Zenodo}~\cite{perez-salinas2023ICdataset}.

\section*{Code availability}
All packages used in this work are open source and available via git repositories online.

\section*{Acknowledgments}
The authors would like to thank Carlo Beenakker, Vedran Dunjko, and Jordi Tura for their support in this project and Patrick Emonts and Yaroslav Herasymenko for useful feedback on the manuscript.
The authors would also like to thank all aQa members for fruitful discussions.
APS acknowledges support from `Quantum Inspire – the Dutch Quantum Computer in the Cloud' project (with project number [NWA.1292.19.194]) of the NWA research program `Research on Routes by Consortia (ORC)', which is funded by the Netherlands Organization for Scientific Research (NWO).
XBM acknowledges funding from the Quantum Software Consortium.

\section*{Competing interests}
The authors declare no competing interests.

\section*{Author contribution}
The project was conceived by XBM and HW.
APS and HW proved the theorems, corollaries, and lemmas.
XBM performed the numerical simulations with assistance from APS and HW.
XBM analyzed the data and performed the model fits.
All the authors interpreted the results and wrote the manuscript.

\bibliography{references} 

\onecolumngrid
\clearpage

\appendix

\section{Proofs}\label{app:proofs}

\subsection{Proof of~\Cref{le:beta}}\label{app:beta}
The main assumption of~\Cref{le:beta} is $\vec{\delta}$ is drawn from the isotropic distribution on the unit sphere in $m$ dimensions.
As a first step, we use the spherical symmetry of the parameter space to align the first coordinate of $\vec\delta$ with the vector $\nabla C(\vec\theta)$. Thus, 
\begin{equation}
    \left(\nabla C(\vtheta) \cdot\vec{\delta}\right)^2  = \norm{\nabla C(\vtheta)}^2\delta_1^2.
\end{equation}
Now, we redefine the isotropic distribution as the normalized multi-dimensional Gaussian distribution (\(\mathcal N\)), 
\begin{equation}
    \vec\delta = \frac{\vec x}{\norm{\vec x}};\qquad {\rm with} \;  \vec x \sim \mathcal{N}(0, \mathbb I^m).
\end{equation}
By definition, each of the coordinates-squared in the multi-dimensional Gaussian distribution follows a $\chi^2$ distribution~\cite{johnson1995continuous}. In particular, $x_1^2 \sim \chi^2(1)$, and 
\(\sum_{i=2}^m x_i^2 \sim \chi^2(m - 1)\). It is well-known~\cite{bailey1992distributional} that the above quotient follows a beta distribution with parameters $1/2$ and $(m-1)/2$, i.e., 
%%%%
\begin{equation}
    \left(\nabla C(\vtheta) \cdot\vec{\delta}\right)^2 = \norm{\nabla C(\vtheta)}^2\frac{x_1^2}{x_1^2 + \sum_{i=2}^m x_i^2} \sim \norm{\nabla C(\vtheta)}^2\Beta\left(\frac{1}{2}, \frac{m-1}{2}\right),
\end{equation}
%%%%
finishing the proof. $\square$

\subsection{Proof of~\Cref{th:cdf}}\label{app:cdf}
The CDF of a beta distribution is the regularized incomplete beta function $\I$. Thus, in the assumptions of~\Cref{le:beta},
%%%%%%
\begin{equation}
    \Pr(\left(\nabla C(\vtheta)\cdot \vec{\delta}\right)^2 \leq \epsilon^2) =\I\left(\frac{\epsilon^2}{||\nabla C(\vtheta)||^2}, \frac{1}{2}, \frac{m-1}{2}\right),
\end{equation}
where  $\epsilon$ is a realization of $(\nabla C(\vtheta) \cdot\vec{\delta})^2$.

We are however interested in $\nabla C(\vec\theta)\cdot\vec\delta$. From the isotropic condition of $\vec\delta$, it is immediate that $\nabla C(\vec\theta)\cdot\vec\delta$ is symmetric with respect to $0$. Using this observation, 
%%%%%%
\begin{align}
    \Pr(|\nabla C(\vtheta)\cdot \vec{\delta}| \leq \epsilon) &= \Pr(-\epsilon \leq\nabla C(\vtheta)\cdot \vec{\delta} \leq 0) + \Pr(0 \leq \nabla C(\vtheta)\cdot \vec{\delta} \leq \epsilon) \\
    &=2\Pr(-\epsilon \leq\nabla C(\vtheta)\cdot \vec{\delta} \leq 0) = 2\Pr(0 \leq \nabla C(\vtheta)\cdot \vec{\delta} \leq \epsilon)\label{eq:individuals}\\
    &= \Pr(\left(\nabla C(\vtheta)\cdot \vec{\delta}\right)^2 \leq \epsilon^2) = \I\left(\frac{\epsilon^2}{||\nabla C(\vtheta)||^2}, \frac{1}{2}, \frac{m-1}{2}\right).
\end{align}
%%%%%%
The step taken in~\Cref{eq:individuals} allows us to rewrite them as
%%%%%
\begin{equation}
\Pr(\nabla C(\vtheta)\cdot \vec{\delta} \leq \epsilon) = \frac{1}{2}\left(1+ \operatorname{sgn}(\epsilon) I\left(\frac{\epsilon^2}{||\nabla C(\vtheta)||^2}, \frac{1}{2}, \frac{m-1}{2}\right)\right),
\end{equation}
%%%%%
where $\operatorname{sgn}$ is the sign function. $\square$

\subsection{Proof of~\Cref{le:h_max}}\label{app:h_max}
For this proof, we must focus on the regime of large values of the IC. We recall the definition of the IC from~\Cref{def:ic}
\begin{equation}
    H = \sum_{a \neq b} h(p_{ab}), 
\end{equation}
with $h(x) = -x\log_6 x$. We define the inverse function $h^{-1}$ to be applied in the domain $x \leq 1/e$. 

For a given value of the sum of probabilities, the maximum entropy is achieved for uniform distribution. This leads to the expression 
\begin{equation}
    \sum_{a \neq b} p_{ab} \geq 6 h^{-1}(H / 6).
\end{equation}
Note that a given value of $H$ is compatible with probability distributions with larger joint probability but uneven distributions. Completeness of the probability distribution implies $\sum_{a \neq b} p_{ab} \leq 1$. The properties of the function $h(x)$ allow maintaining a value $H$, with one probability $p_1$ to decrease for a given quantity $x$, as long as another probability increases some other quantity $f(x) > x$.  Hence, a high value of $H$ implies a minimal value on at least some set of probabilities. 

We focus on the probability held by only 4 elements in the probability distribution. We first split the IC value into two pieces, the 4 smallest ones and the 2 largest, 
\begin{equation}
    H = \sum_{4} h(p_{ab}) + \sum_{2} h(p_{ab}),
\end{equation}
where $\sum_{4}$ indicates the sum over the smallest terms, and $\sum_2$ stands for the largest terms. To obtain the minimal probability held by the smallest 4 terms, we start in the situation with the smallest possible sum of all 6 probabilities, namely $p_{ab} = h^{-1}(H / 6) \equiv q, \forall a, b$. Now we subtract probability from these 4 terms and add probability to the remaining terms to keep the IC constant. 
\begin{equation}
    H = 4 h(q - x_1 - x_2) + h(q + f_1(x_1)) + h(q + f_2(x_2)),
\end{equation} 
where $f_{1, 2}$ is whatever function needed. Concavity of the function $h$ allows us to bound
\begin{equation}\label{eq:boundH}
    H \leq 4 h(q - x) + 2 h(q + f(x)). 
\end{equation}
This bound holds as long as
\begin{equation}
    4(q - x) + 2(q + f(x)) \leq 1, 
\end{equation}
where the equality is satisfied under the limit case. Substituting this equality into~\Cref{eq:boundH}, we obtain 
\begin{equation}
    H \leq 4 h(q - x) + 2 h\left( \frac{1}{2} - 2 q + 2 x\right), 
\end{equation}
which comprises the values of $x$ compatible with $H$. This bound also considers probability transferred to elements not relevant to the IC. Recalling $p_4$, we know 
\begin{equation}
     \sum_{4} p_{ab} \geq 4(q - x), 
\end{equation}
and a more straightforward version of this condition is written as
\begin{equation}
    \sum_{4} p_{ab} \geq 4 q_4, 
\end{equation}
with $q_4$ being the solution to the equation $H = 4 h(x) + 2 h(1/2 - 2x)$. $\square$

\subsection{Proof of~\Cref{le:h_s}}\label{app:h_s}
The first step is to observe that in case of sufficiently small IC, there are only two possible scenarios, which are a) only one of the probabilities $p_{ab}$, with $a \neq b$ is close to one, and b) all $p_{ab}$ are close to zero. This scenario a) is impossible by construction since these probabilities must come at least in pairs. Thus, all of them are small. In particular, since $h(x) \geq x$ for $x \leq 1 / 6$, we bound
\begin{equation}
    \sum_{a \neq b} p_{ab} \leq H \leq \eta \leq 1/6.
\end{equation}
Since all probabilities are small and are also combinations of the probabilities of only one step to be $p_+, p_-, p_\odot$, we can conclude that at least two of those must be small. Since $p_\pm$ are symmetric by construction, the only candidate left is $p_{\odot}$, which is the event that concentrates all probabilities. This observation allows us to bound 
\begin{equation}
    p_{++} \leq p_{\odot +} \leq \eta, 
\end{equation}
and subsequently
\begin{equation}
    p_{\odot\odot} = 1 - p_{++} - p_{--} - \sum_{a \neq b} p_{ab} \geq 1 - 3 \eta.
\end{equation}
$\square$

\subsection{Proof of~\Cref{th:h_m_grad}}\label{app:h_m_grad}
For this theorem, we need to bound the probability of one step of the random walk to be $\{-, \odot, +\}$. However, the IC measures only pairs of steps. In this proof, we use the results from~\Cref{le:h_max} to bound probabilities in only one step, and we connect those to the results in~\Cref{cor:cdf}. We take, without loss of generality $p_+$, 
\begin{equation}
    p_{+} = p_{+ +} + \frac{1}{2}\left( p_{+\odot} + p_{\odot +} + p_{- +} + p_{+ -} \right)
\end{equation}
The IC is insensitive to $p_{++}$, so we discard it. The second term is bounded by~\Cref{le:h_max}, thus
\begin{equation}
    p_+ \geq 2 q_4, 
\end{equation}
with $q_4$ being the solution to the equation $H = 4 h(x) + 2 h(1/2 - 2x)$. Now, we recall~\Cref{cor:cdf} and bound
\begin{equation}
    \Phi_{G}\left(\frac{\epsilon {\sqrt m}}{\Vert \nabla C \Vert_W} \right) \geq 2 q_4 , 
\end{equation}
which directly leads to
\begin{equation}
    \Normgrad_W \geq \frac{- \epsilon {\sqrt{m}}}{\Phi_{G}^{-1}(2 q_4)},
\end{equation}
yielding the desired result, up to a symmetry of the CDF function. The upper bound can be obtained by applying the same reasoning to $p_0 \geq 2q_4$. $\square$

\subsection{Proof of~\Cref{th:h_s_grad}}\label{app:h_s_grad}
For this theorem, we need to bound the probability of one step of the random walk to be $\odot$. The results from~\Cref{le:h_s} bound $P_{\odot\odot}$, so
\begin{equation}
    p_\odot \geq p_{\odot\odot} \geq 1 - 3\eta
\end{equation}
Now we connect this bound to the results in~\Cref{cor:cdf}. 
\begin{equation}
    \Phi_m\left(\frac{-\epsilon}{\Normgrad_W} \right) \leq \frac{3}{2}\eta,
\end{equation}
which yields the result
\begin{equation}
    \Normgrad_W \leq \frac{-\epsilon}{\Phi_m^{-1}(3\eta/2)}.
\end{equation}
$\square$

\section{Details on the numerical experiments}\label{app:numerical_experiments}
The numerical experiments in \Cref{sec:numerics} have been done using the quantum simulator software \textsc{Qibo}~\cite{efthymiou2022qibo} together  with \textsc{Flacco}~\cite{kerschke2019comprehensive} to compute the information content analysis.
For each number of qubits and layers, we perform 5 independent repetitions.
The results are then the median of these runs, and the error bars on the figures depict their standard deviations.
The data and code to reproduce the results of this paper can be found in~\cite{perez-salinas2023ICdataset}.

The process to generate the dataset is as follows:
\begin{enumerate}
    \item Call \textsc{Flacco} to generate a Latin-hyper cube sampling of the parameter space.
    \item Use \textsc{Qibo} to compute \(C(\vtheta)\).
    \item Compute \(H_{M}\), \(\epsM\), and \(\epsS\) from the randomly sampled tuples \(\big[\vtheta_i, C(\vtheta_i)\big]\) with \href{https://cran.r-project.org/web/packages/flacco/index.html}{\textsc{Flacco}}.
\end{enumerate}

\subsection{Scaling pre-factors results}
To estimate the scaling pre-factors, we conducted curve-fitting with the software \textsc{StatsModels}~\cite{seabold2010statsmodels}, using the ordinary least-squares method.

The software provides a summary of the results of the fit with interesting information about its quality.
We have not provided any details about the fit quality to avoid a flood of data but the summaries can be found together with the dataset in \cite{perez-salinas2023ICdataset}.

In this appendix, we extend \Cref{tab:global_scaling_qubits}, \Cref{tab:local_scaling_qubits}, and \Cref{tab:local_scaling_layers} to include the upper bounds on the coefficients of the fit.

\begin{figure}
\centering
    \includegraphics[width=.7\textwidth]{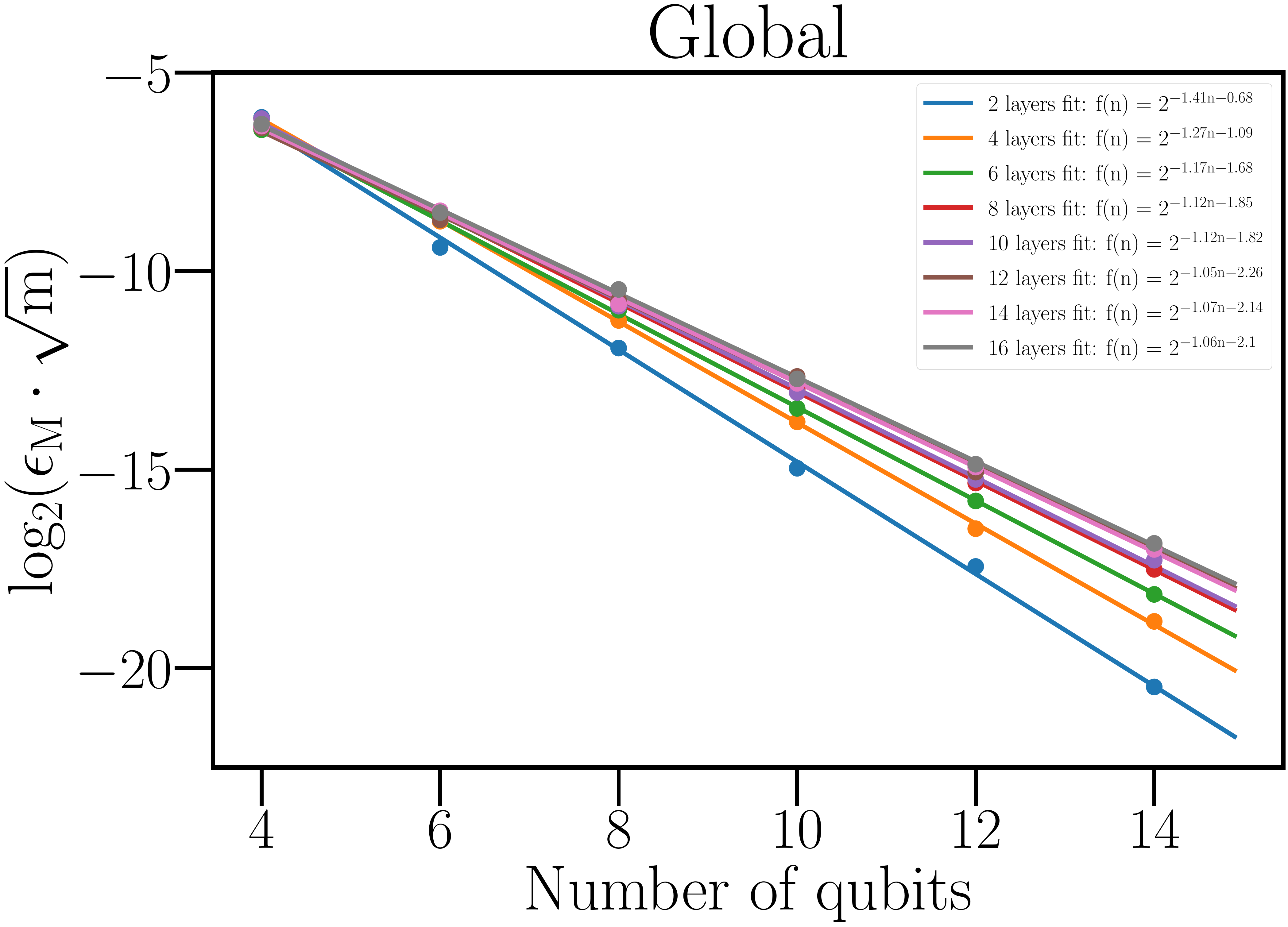}
    \caption{Linear fit of the \(\rm{log}_2(\epsM \cdot \sqrt{m})\) with respect to qubits for the global cost function.}
    \label{fig:app_global_scaling_fit}
\end{figure}

\begin{figure}
\subfloat{
    \centering
    \includegraphics[width=.425\textwidth]{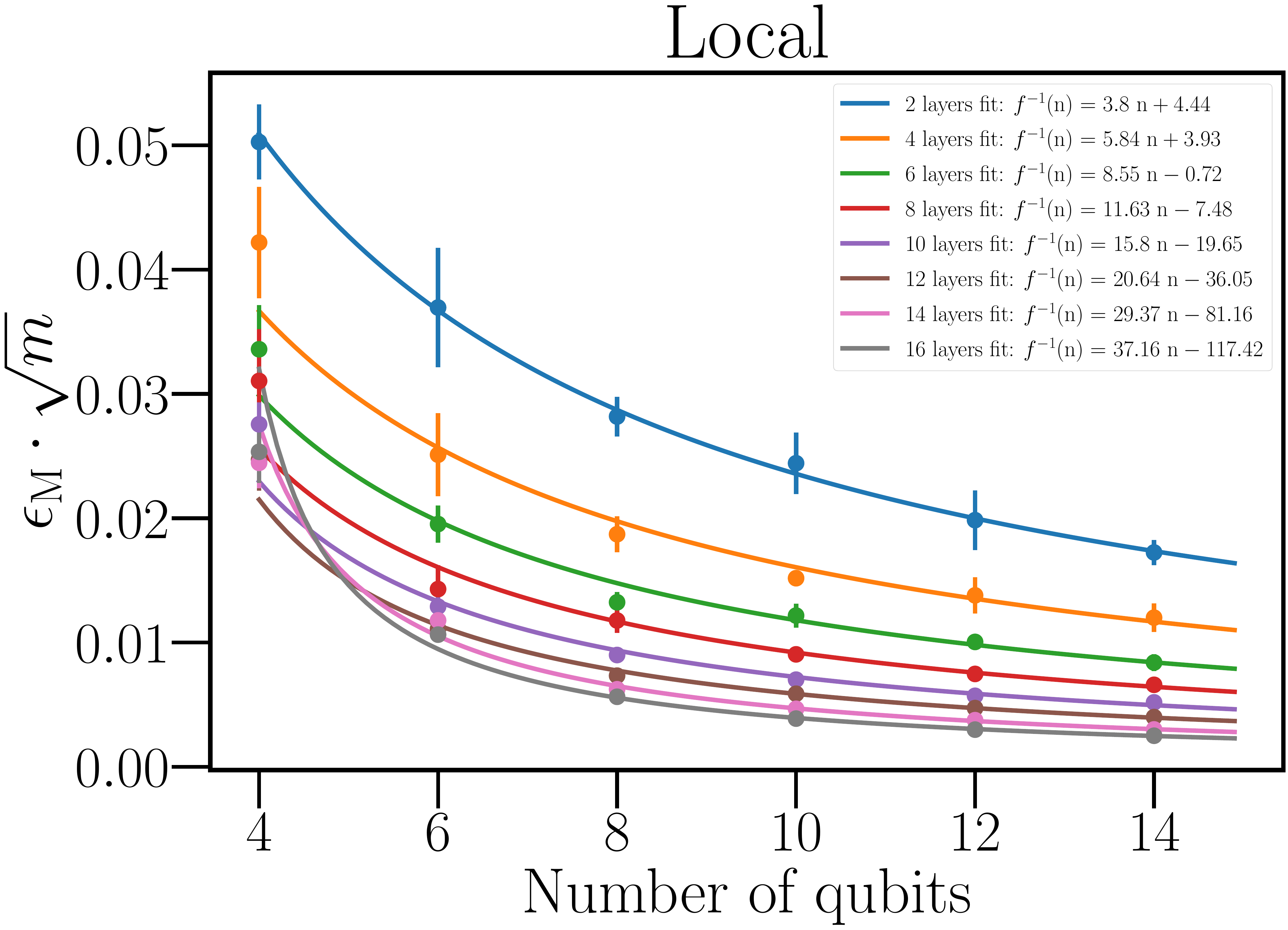}
}
\subfloat{
    \centering
    \includegraphics[width=.425\textwidth]{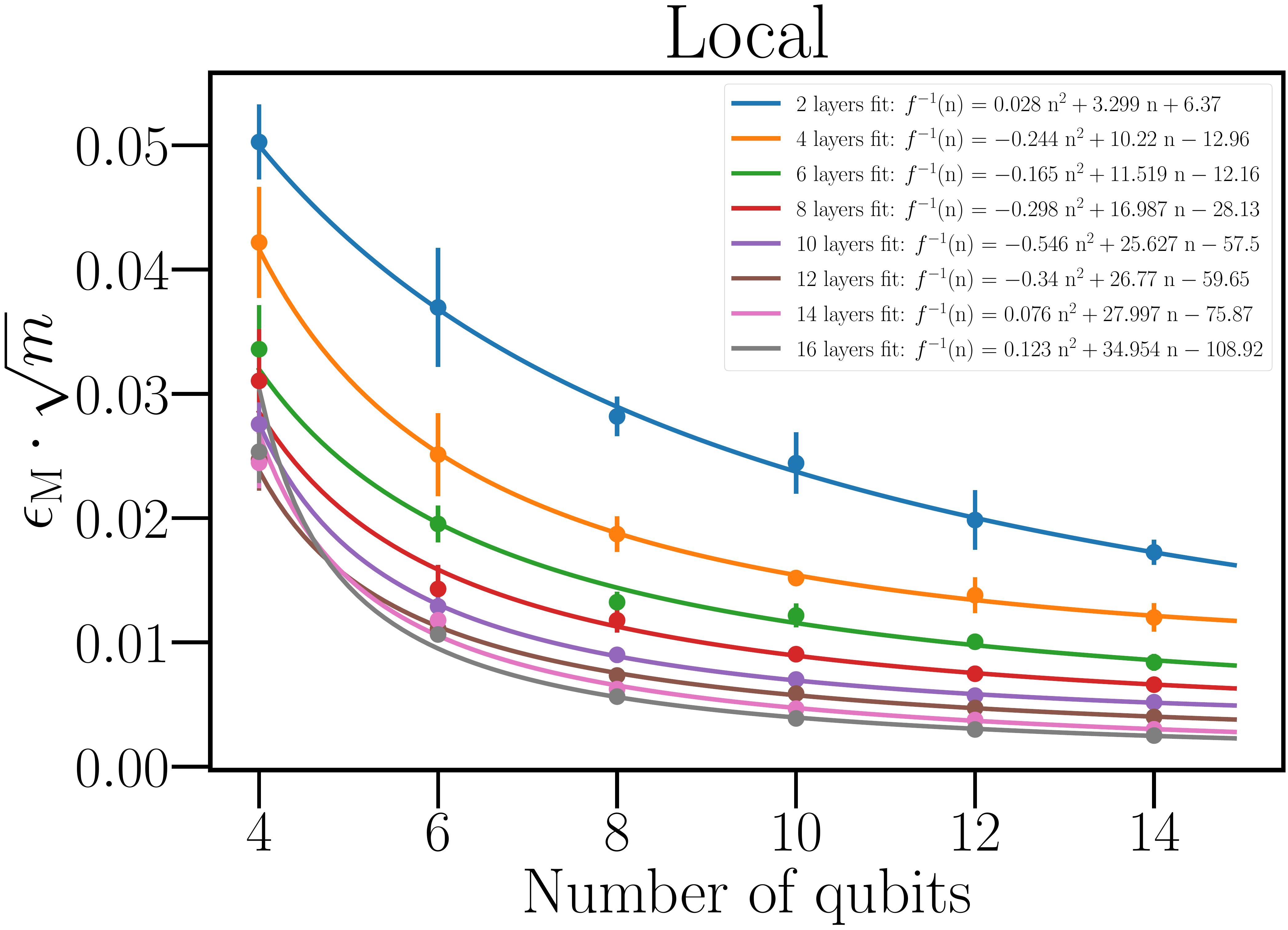}
}
\caption{Pre-factor scaling fit for the local cost function with the number of qubits. The left panel depicts the linear model results. The right panel corresponds to the quadratic polynomial model.
The error bars on the figures are the standard deviation of a sample of 5 independent runs.}
\label{fig:app_local_qubit_scaling_fit}
\end{figure}

\begin{figure}
\subfloat{
    \centering
    \includegraphics[width=.425\textwidth]{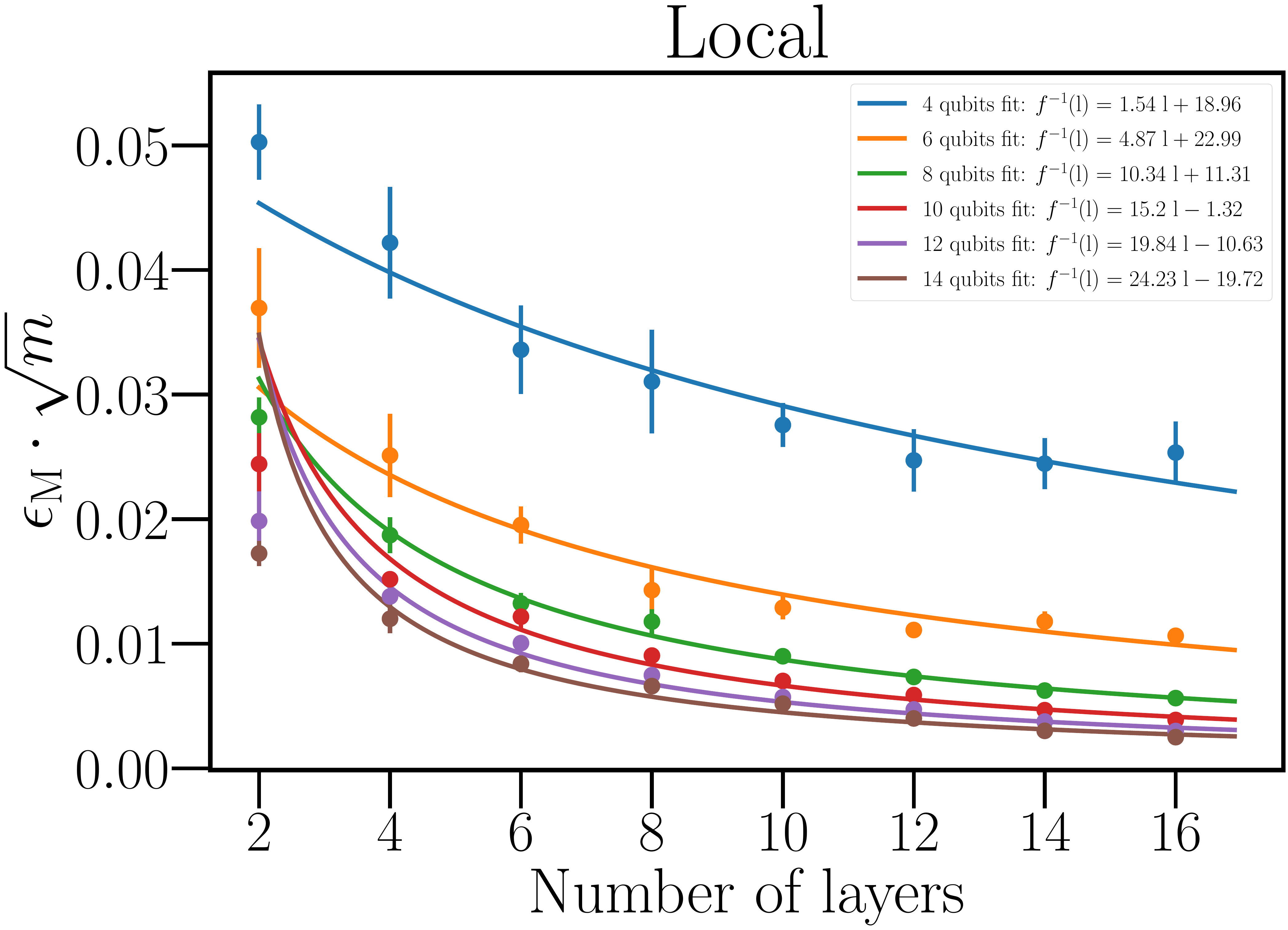}
}
\subfloat{
    \centering
    \includegraphics[width=.425\textwidth]{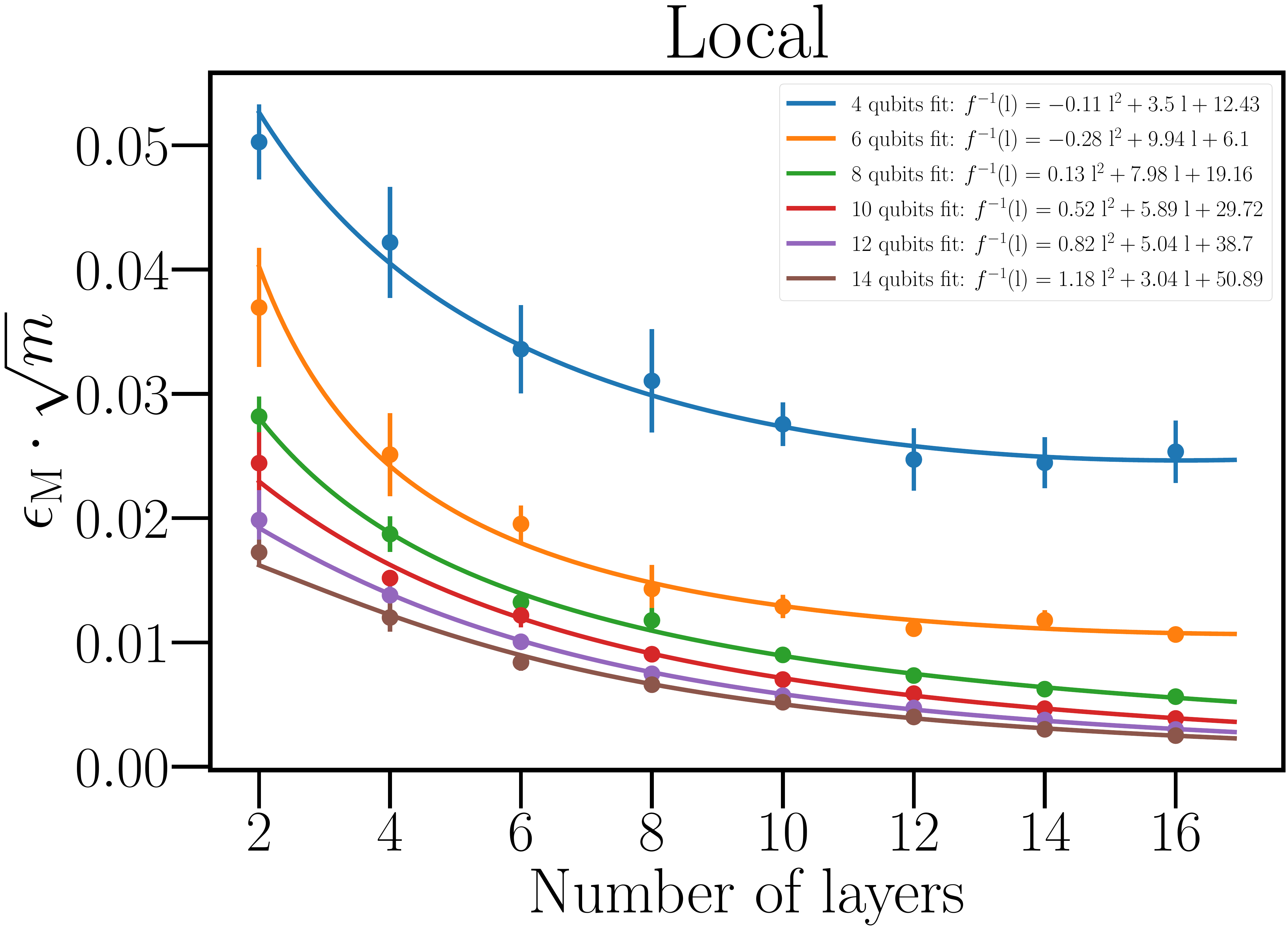}
}
\caption{Pre-factor scaling fit for the local cost function with the number of layers. The left panel depicts the linear model results. The right panel corresponds to the quadratic polynomial model.
The error bars on the figures are the standard deviation of a sample of 5 independent runs.}
\label{fig:app_local_layer_scaling_fit}
\end{figure}

We also include figures to show the quality of the fits.
\Cref{fig:app_global_scaling_fit} shows the fit of the global cost function to a linear model.
In \Cref{fig:app_local_qubit_scaling_fit} and \Cref{fig:app_local_layer_scaling_fit} we show the fits for the local cost function scaling with qubits and layers respectively.
We fit a linear model (left panel) and degree-two polynomial (right panel) to visualize and compare the fit qualities.

\begin{table*}[t]
\begin{tabular}{|ccccccccc|}
\hline
\multicolumn{9}{|c|}{Global cost function scaling with qubits}                                                                                                                                                                                         \\ \hline
\multicolumn{1}{|c|}{\(f(n) = 2^{\alpha n + \beta}\)}  & \multicolumn{4}{c|}{\(\alpha\)} & \multicolumn{4}{c|}{\(\beta\)}      \\ \hline\hline
\multicolumn{1}{|c|}{Layers} & \multicolumn{1}{c|}{LB}    & \multicolumn{1}{c|}{\(\epsMp\)} & \multicolumn{1}{c|}{UB} & \multicolumn{1}{c|}{UBs} & \multicolumn{1}{c|}{LB}   & \multicolumn{1}{c|}{\(\epsMp\)} & \multicolumn{1}{c|}{UB} & UBs \\ \hline
\multicolumn{1}{|c|}{2}      & \multicolumn{1}{c|}{-1.43}      & \multicolumn{1}{c|}{{-1.41}}     & \multicolumn{1}{c|}{-1.37}   & \multicolumn{1}{c|}{-0.74}    & \multicolumn{1}{c|}{-0.91}     & \multicolumn{1}{c|}{{-0.68}}     & \multicolumn{1}{c|}{1.87}   &   1.12  \\ \hline
\multicolumn{1}{|c|}{4}      & \multicolumn{1}{c|}{-1.29}      & \multicolumn{1}{c|}{{-1.27}}     & \multicolumn{1}{c|}{-1.23}   & \multicolumn{1}{c|}{-0.77}    & \multicolumn{1}{c|}{-1.22}     & \multicolumn{1}{c|}{{-1.09}}     & \multicolumn{1}{c|}{1.34}   &   0.93  \\ \hline
\multicolumn{1}{|c|}{6}      & \multicolumn{1}{c|}{-1.19}      & \multicolumn{1}{c|}{{-1.17}}     & \multicolumn{1}{c|}{-1.15}   & \multicolumn{1}{c|}{-0.77}    & \multicolumn{1}{c|}{-1.85}     & \multicolumn{1}{c|}{{-1.68}}    & \multicolumn{1}{c|}{0.84}   &   0.44  \\ \hline
\multicolumn{1}{|c|}{8}      & \multicolumn{1}{c|}{-1.13}      & \multicolumn{1}{c|}{{-1.12}}     & \multicolumn{1}{c|}{-1.10}   & \multicolumn{1}{c|}{-0.80}    & \multicolumn{1}{c|}{-1.98}     & \multicolumn{1}{c|}{{-1.85}}     & \multicolumn{1}{c|}{0.64}   &   0.37  \\ \hline
\multicolumn{1}{|c|}{10}     & \multicolumn{1}{c|}{-1.12}      & \multicolumn{1}{c|}{{-1.12}}     & \multicolumn{1}{c|}{-1.10}   & \multicolumn{1}{c|}{-0.85}    & \multicolumn{1}{c|}{-1.97}     & \multicolumn{1}{c|}{{-1.82}}     & \multicolumn{1}{c|}{0.73}   &   0.49  \\ \hline
\multicolumn{1}{|c|}{12}     & \multicolumn{1}{c|}{-1.06}      & \multicolumn{1}{c|}{{-1.05}}     & \multicolumn{1}{c|}{-1.04}   & \multicolumn{1}{c|}{-0.88}    & \multicolumn{1}{c|}{-2.41}     & \multicolumn{1}{c|}{{-2.26}}     & \multicolumn{1}{c|}{0.27}   &   0.59  \\ \hline
\multicolumn{1}{|c|}{14}     & \multicolumn{1}{c|}{-1.07}      & \multicolumn{1}{c|}{{-1.07}}     & \multicolumn{1}{c|}{-1.06}   & \multicolumn{1}{c|}{-0.89}    & \multicolumn{1}{c|}{-2.29}     & \multicolumn{1}{c|}{{-2.14}}     & \multicolumn{1}{c|}{0.40}   &   0.50  \\ \hline
\multicolumn{1}{|c|}{16}     & \multicolumn{1}{c|}{-1.06}      & \multicolumn{1}{c|}{{-1.06}}     & \multicolumn{1}{c|}{-1.05}   & \multicolumn{1}{c|}{-0.91}    & \multicolumn{1}{c|}{-2.25}     & \multicolumn{1}{c|}{{-2.10}}     & \multicolumn{1}{c|}{0.45}   &   0.64  \\ \hline
\end{tabular}
\caption{Extended results on estimating the coefficients for the global cost function scaling with qubits.
The additional columns show the upper bound from $\epsM$ (UB) and upper bound from $\epsS$ (UBs).}
\label{tab:app_global_scaling_qubits}
\end{table*}

\begin{table*}[]
\begin{tabular}{|ccccccccccccc|}
\hline
\multicolumn{13}{|c|}{Local cost function scaling with qubits}                                                                                                                                                                                                                                                                                               \\ \hline
\multicolumn{1}{|c|}{\(f(n) = \alpha n^{2} + \beta n + \gamma\)}   & \multicolumn{4}{c|}{\(\alpha\)}                                                                               & \multicolumn{4}{c|}{\(\beta\)}                                                                                & \multicolumn{4}{c|}{\(\gamma\)}                                                          \\ \hline\hline
\multicolumn{1}{|c|}{Layers} & \multicolumn{1}{c|}{LB} & \multicolumn{1}{c|}{\(\epsMp\)} & \multicolumn{1}{c|}{UB} & \multicolumn{1}{c|}{UBs} & \multicolumn{1}{c|}{LB} & \multicolumn{1}{c|}{\(\epsMp\)} & \multicolumn{1}{c|}{UB} & \multicolumn{1}{c|}{UBs} & \multicolumn{1}{c|}{LB} & \multicolumn{1}{c|}{\(\epsMp\)} & \multicolumn{1}{c|}{UB} & UBs \\ \hline

\multicolumn{1}{|c|}{2}      & \multicolumn{1}{c|}{0.05}   & \multicolumn{1}{c|}{0.03}     & \multicolumn{1}{c|}{0}   & \multicolumn{1}{c|}{0.01}    & \multicolumn{1}{c|}{2.80}   & \multicolumn{1}{c|}{3.3}     & \multicolumn{1}{c|}{0.79}   & \multicolumn{1}{c|}{0.87}    & \multicolumn{1}{c|}{8.05}   & \multicolumn{1}{c|}{6.37}     & \multicolumn{1}{c|}{1.23}   &    1.49 \\ \hline
\multicolumn{1}{|c|}{4}      & \multicolumn{1}{c|}{-0.25}   & \multicolumn{1}{c|}{-0.24}     & \multicolumn{1}{c|}{-0.05}   & \multicolumn{1}{c|}{0}    & \multicolumn{1}{c|}{10.08}   & \multicolumn{1}{c|}{10.22}     & \multicolumn{1}{c|}{2.09}   & \multicolumn{1}{c|}{1.81}    & \multicolumn{1}{c|}{-12.86}   & \multicolumn{1}{c|}{-12.96}     & \multicolumn{1}{c|}{-2.37}   &  -0.50   \\ \hline
\multicolumn{1}{|c|}{6}      & \multicolumn{1}{c|}{-0.16}   & \multicolumn{1}{c|}{-0.16}     & \multicolumn{1}{c|}{-0.04}   & \multicolumn{1}{c|}{-0.08}    & \multicolumn{1}{c|}{11.09}   & \multicolumn{1}{c|}{11.52}     & \multicolumn{1}{c|}{2.45}   & \multicolumn{1}{c|}{3.91}    & \multicolumn{1}{c|}{-11.37}   & \multicolumn{1}{c|}{-12.16}     & \multicolumn{1}{c|}{-2.46}   &  -5.73  \\ \hline
\multicolumn{1}{|c|}{8}      & \multicolumn{1}{c|}{-0.27}   & \multicolumn{1}{c|}{-0.30}     & \multicolumn{1}{c|}{-0.07}   & \multicolumn{1}{c|}{-0.12}    & \multicolumn{1}{c|}{16.20}   & \multicolumn{1}{c|}{16.99}     & \multicolumn{1}{c|}{3.67}   & \multicolumn{1}{c|}{5.59}    & \multicolumn{1}{c|}{-26.22}   & \multicolumn{1}{c|}{-28.13}     & \multicolumn{1}{c|}{-6.09}   &   -11.48  \\ \hline
\multicolumn{1}{|c|}{10}     & \multicolumn{1}{c|}{-0.56}   & \multicolumn{1}{c|}{-0.55}     & \multicolumn{1}{c|}{-0.11}   & \multicolumn{1}{c|}{-0.08}    & \multicolumn{1}{c|}{25.33}   & \multicolumn{1}{c|}{25.63}     & \multicolumn{1}{c|}{5.25}   & \multicolumn{1}{c|}{6.49}    & \multicolumn{1}{c|}{-57.30}   & \multicolumn{1}{c|}{-57.50}     & \multicolumn{1}{c|}{-11.4}   &  -14.87   \\ \hline
\multicolumn{1}{|c|}{12}     & \multicolumn{1}{c|}{-0.31}   & \multicolumn{1}{c|}{-0.34}     & \multicolumn{1}{c|}{-0.08}   & \multicolumn{1}{c|}{-0.05}    & \multicolumn{1}{c|}{25.52}   & \multicolumn{1}{c|}{26.77}     & \multicolumn{1}{c|}{5.79}   & \multicolumn{1}{c|}{7.41}    & \multicolumn{1}{c|}{-56.33}   & \multicolumn{1}{c|}{-59.65}     & \multicolumn{1}{c|}{-12.93}   &  -18.09   \\ \hline
\multicolumn{1}{|c|}{14}     & \multicolumn{1}{c|}{0.14}   & \multicolumn{1}{c|}{0.08}     & \multicolumn{1}{c|}{-0.01}   & \multicolumn{1}{c|}{-0.12}    & \multicolumn{1}{c|}{25.99}   & \multicolumn{1}{c|}{28.00}     & \multicolumn{1}{c|}{6.28}   & \multicolumn{1}{c|}{10.42}    & \multicolumn{1}{c|}{-69.42}   & \multicolumn{1}{c|}{-75.87}     & \multicolumn{1}{c|}{-17.24}   &   -30.12  \\ \hline
\multicolumn{1}{|c|}{16}     & \multicolumn{1}{c|}{0.13}   & \multicolumn{1}{c|}{0.12}     & \multicolumn{1}{c|}{0.02}   & \multicolumn{1}{c|}{-0.10}    & \multicolumn{1}{c|}{33.69}   & \multicolumn{1}{c|}{34.95}     & \multicolumn{1}{c|}{7.45}   & \multicolumn{1}{c|}{12.28}    & \multicolumn{1}{c|}{-104.77}   & \multicolumn{1}{c|}{-108.92}     & \multicolumn{1}{c|}{-23.17}   &  -38.98   \\ \hline
\end{tabular}
\caption{Extended results on estimating the coefficients for the local cost function scaling with qubits.
The additional columns show the upper bound from $\epsM$ (UB) and upper bound from $\epsS$ (UBs).}
\label{tab:app_local_scaling_qubits}
\end{table*}

\begin{table*}[]
\begin{tabular}{|ccccccccccccc|}
\hline
\multicolumn{13}{|c|}{Local cost function scaling with layers}                                                                                                                                                                                                                                                                                               \\ \hline
\multicolumn{1}{|c|}{\(f(l) = \alpha l^{2} + \beta l + \gamma\)}   & \multicolumn{4}{c|}{\(\alpha\)}                                                                               & \multicolumn{4}{c|}{\(\beta\)}                                                                                & \multicolumn{4}{c|}{\(\gamma\)}                                                          \\ \hline\hline
\multicolumn{1}{|c|}{Qubits} & \multicolumn{1}{c|}{LB} & \multicolumn{1}{c|}{\(\epsMp\)} & \multicolumn{1}{c|}{UB} & \multicolumn{1}{c|}{UBs} & \multicolumn{1}{c|}{LB} & \multicolumn{1}{c|}{\(\epsMp\)} & \multicolumn{1}{c|}{UB} & \multicolumn{1}{c|}{UBs} & \multicolumn{1}{c|}{LB} & \multicolumn{1}{c|}{\(\epsMp\)} & \multicolumn{1}{c|}{UB} & UBs \\ \hline

\multicolumn{1}{|c|}{4}      & \multicolumn{1}{c|}{-0.1}   & \multicolumn{1}{c|}{-0.11}     & \multicolumn{1}{c|}{-0.02}   & \multicolumn{1}{c|}{-0.03}    & \multicolumn{1}{c|}{3.29}   & \multicolumn{1}{c|}{3.5}     & \multicolumn{1}{c|}{0.73}   & \multicolumn{1}{c|}{0.97}    & \multicolumn{1}{c|}{12.74}   & \multicolumn{1}{c|}{12.43}     & \multicolumn{1}{c|}{2.82}   &   3.32  \\ \hline
\multicolumn{1}{|c|}{6}      & \multicolumn{1}{c|}{-0.27}   & \multicolumn{1}{c|}{-0.28}     & \multicolumn{1}{c|}{-0.06}   & \multicolumn{1}{c|}{-0.04}    & \multicolumn{1}{c|}{9.54}   & \multicolumn{1}{c|}{9.94}     & \multicolumn{1}{c|}{2.10}   & \multicolumn{1}{c|}{2.15}    & \multicolumn{1}{c|}{6.26}   & \multicolumn{1}{c|}{6.10}     & \multicolumn{1}{c|}{1.51}   &    3.14 \\ \hline
\multicolumn{1}{|c|}{8}      & \multicolumn{1}{c|}{0.13}   & \multicolumn{1}{c|}{0.13}     & \multicolumn{1}{c|}{0.03}   & \multicolumn{1}{c|}{0.01}    & \multicolumn{1}{c|}{7.69}   & \multicolumn{1}{c|}{7.98}     & \multicolumn{1}{c|}{1.67}   & \multicolumn{1}{c|}{2.85}    & \multicolumn{1}{c|}{18.80}   & \multicolumn{1}{c|}{19.16}     & \multicolumn{1}{c|}{4.29}   &   2.82  \\ \hline
\multicolumn{1}{|c|}{10}     & \multicolumn{1}{c|}{0.48}   & \multicolumn{1}{c|}{0.52}     & \multicolumn{1}{c|}{0.12}   & \multicolumn{1}{c|}{0.14}    & \multicolumn{1}{c|}{5.98}   & \multicolumn{1}{c|}{5.89}     & \multicolumn{1}{c|}{1.13}   & \multicolumn{1}{c|}{2.08}    & \multicolumn{1}{c|}{27.79}   & \multicolumn{1}{c|}{29.72}     & \multicolumn{1}{c|}{6.88}   &   7.27  \\ \hline
\multicolumn{1}{|c|}{12}     & \multicolumn{1}{c|}{0.81}   & \multicolumn{1}{c|}{0.82}     & \multicolumn{1}{c|}{0.17}   & \multicolumn{1}{c|}{0.26}    & \multicolumn{1}{c|}{4.68}   & \multicolumn{1}{c|}{5.04}     & \multicolumn{1}{c|}{1.10}   & \multicolumn{1}{c|}{1.34}    & \multicolumn{1}{c|}{38.24}   & \multicolumn{1}{c|}{38.70}     & \multicolumn{1}{c|}{8.19}   &   11.04  \\ \hline
\multicolumn{1}{|c|}{14}     & \multicolumn{1}{c|}{1.14}   & \multicolumn{1}{c|}{1.18}     & \multicolumn{1}{c|}{0.25}   & \multicolumn{1}{c|}{0.24}    & \multicolumn{1}{c|}{2.95}   & \multicolumn{1}{c|}{3.04}     & \multicolumn{1}{c|}{0.61}   & \multicolumn{1}{c|}{2.50}    & \multicolumn{1}{c|}{48.93}   & \multicolumn{1}{c|}{50.89}     & \multicolumn{1}{c|}{11.11}   &  10.36   \\ \hline
\end{tabular}
\caption{Extended results on estimating the coefficients for the local cost function scaling with layers.
The additional columns show the upper bound from $\epsM$ (UB) and upper bound from $\epsS$ (UBs).}
\label{tab:app_local_scaling_layers}
\end{table*}

\end{document}